\documentclass[prb,reprint,twocolumn,showpacs,superscriptaddress]{revtex4-1}

\usepackage{amsmath}

\usepackage{graphicx}
\usepackage{dcolumn}
\usepackage{bm}
\usepackage{epstopdf}
\usepackage{setspace}  
\newcommand{\beq}{\begin{equation}}
\newcommand{\eeq}{\end{equation}}
\newcommand{\ba}{\begin{array}{ccc}}
\newcommand{\ea}{\end{array}}
\newcommand{\nn}{\nonumber}
\newcommand{\T}{{\mathcal T}}
\newcommand{\C}{{\mathcal C}}
\newcommand{\Tt}{\tilde{\mathcal \T}}
 \renewcommand{\d}{\partial}
\def\bea{\begin{eqnarray}}
\def\eea{\end{eqnarray}}

\def\<{\langle}
\def\>{\rangle}

\usepackage{amsmath}
\usepackage{amssymb}
\usepackage{color}
\begin{document}
\title{Particle-vortex duality of 2D Dirac fermion from electric-magnetic duality of 3D topological insulators}

\author{Max A. Metlitski}
\affiliation{Kavli Institute for Theoretical Physics, UC Santa Barbara, CA 93106, USA. }

\author{Ashvin Vishwanath}
\affiliation{Department of Physics, University of California, Berkeley, CA 94720, USA.}
\affiliation{Materials Science Division, Lawrence Berkeley National Laboratories, Berkeley, CA 94720, USA.}

\begin{abstract} Particle-vortex duality is a powerful theoretical tool that has been used to study bosonic systems. Here we propose an analogous duality for Dirac fermions in 2+1 dimensions. The physics of a single Dirac cone is proposed to be described by a dual theory, QED$_3$ with a dual Dirac fermion  coupled to a gauge field.  This duality is established by considering two alternate descriptions of the 3d topological insulator (TI) surface.  The first description is the usual Dirac cone surface state. The second  description is accessed via an electric-magnetic duality of the {  bulk} TI coupled to a gauge field, which maps it  to a gauged topological superconductor. This alternate description ultimately leads to a new surface theory - dual QED$_3$. 
The dual theory provides an explicit derivation  of the T-Pfaffian state, a proposed surface topological order of the TI, which is simply the paired superfluid state of the dual fermions. The roles of time reversal and particle-hole symmetry are exchanged by the duality, which connects some of our results to a recent conjecture by Son on  particle-hole symmetric quantum Hall states.

\end{abstract}

\maketitle
\section{Introduction} 
\label{sec:Intro}
Following the prediction and classification of topological insulators (TIs) and topological superconductors (TSc) based on free fermion models \cite{FranzMolenkamp2013}, the conceptual frontier has now shifted to studying strongly interacting topological phases.  For low dimensional phases, one can utilize powerful non-perturbative techniques to describe the 1+1D edge or bulk to obtain qualitatively new physics introduced by interactions \cite{Haldane,Chen1d, Fidkowski0, Fidkowski1,Turner1d,LevinGu,LuAV_SPT}. However, for 3+1D systems, we have few non-perturbative tools - nevertheless remarkable theoretical progress has been made in recent years. For example, entirely new phases that only appear in interacting systems have been predicted \cite{AVTS, MetlitskibTI, Bi2013,Kitaev_pc, Kapustin2014,Kapustin2014a,ChongScience,Senthil2014}.   Furthermore, phases that were {  predicted to be distinct by} the free fermion classification, and whose edge states are stable to weak interactions, can sometimes be smoothly connected in the presence of strong interactions. A striking example of this phenomen {  is provided by} the 3d topological superconductors with time reversal invariance (class DIII), whose integer free fermion classification is broken down to Z$_{16}$ \cite{Kitaev_pc,FidkowskiChenAV,Wang2014,MetlitskiChenFidkowskiAV2014,You2014} by strong interactions. The  discovery of strongly correlated TIs such as 3D topological Kondo insulators \cite{Coleman2010} may provide an experimental window into the effects of strong interactions.

A useful theoretical tool that was introduced to study surfaces of strongly interacting 3d topological phases is surface topological order (STO)  \cite{AVTS,Wang2013,Burnell2013,FidkowskiChenAV,Bonderson2013,Metlitski2013,Wang2013a,Chen2014PRB,Wang2014,MetlitskiChenFidkowskiAV2014}.  In the early days of the field it was assumed that the surface of a topological phase, such as a topological insulator, is metallic if all symmetries are preserved. The resulting surface state of a TI, a single Dirac cone, is forbidden in a purely 2d system with time reversal invariance and charge conservation, since it suffers from the parity anomaly.\cite{Redlich,Semenoff} If gapped, it was generally assumed that the surface must break one of the protecting symmetries such as time reversal symmetry. However, with strong interactions, new possibilities arise. A gapped, insulating surface state of the TI can preserve all symmetries if it is topologically ordered, i.e. if the surface supports anyonic excitations  with fractional quantum numbers. This topological order must encode the parity anomaly that ensures it is a bona fide surface state of the topological insulator bulk. In this sense it encodes the same `Hilbert space', with the same anomalies as the single Dirac cone surface state. Here we will discuss a dual surface theory that also captures the same Hilbert space - which, in contrast to the STO, is gapless  in the UV and is described by QED$_3$. 

More precisely, the surface Dirac theory of a TI is given by the Lagrangian:
\beq
{\mathcal L}_e = \bar{\Psi}_e i \gamma^\mu[\partial_\mu -iA_\mu]\Psi_e
\label{eq:LDirace}
\eeq
where $\bar{\Psi}_e = \Psi^\dagger_e\gamma^0$, $\gamma^\mu$ are $2\times 2$ Dirac matrices,
and we have introduced an external electromagnetic potential $A_{\mu}$ to keep track of the conserved U(1) charge, and possibly insert a chemical potential. Then, the proposed dual surface theory is:
\beq
 {\mathcal L}_{cf} = \bar{\psi}_{cf} i \gamma^\mu[\partial_\mu -i a_\mu]\psi_{cf} - \frac{ 1}{4\pi} \epsilon^{\mu\nu\lambda} A_\mu\partial_\nu a_\lambda
\label{duaL}
\eeq
where the fermions are now coupled to an emergent gauge field $a_{\mu}$, whose flux  is proportional to the electron density, or more precisely, $4\pi$ flux of $a$ corresponds to unit electron charge. 

Let us note three key points. First, we can ask - how do we represent the electron insertion operator $\Psi_e$ in the dual theory?  We find that $\Psi_e$ corresponds to a double monopole operator that introduces $4\pi$ flux as expected from the previous discussion. It will be shown that this operator has all the desired properties. Second - how do we interpret  $\psi_{cf}$, the dual fermions, in terms of electrons? $\psi_{cf}$ will be shown to be a double {  ($2hc/e$)} vortex in the electron fluid bound to an electron - closely analogous to the composite fermion construction \cite{HLR,Jain} - which accounts for the subscript. Finally, we note the action of time reversal symmetry exchanges particles and holes of $\psi_{cf}$, $T:\, \psi_{cf} \rightarrow \psi_{cf}^\dagger$, consistent with their interpretation as vortex like degrees of freedom. A finite chemical potential on the electronic Dirac cone translates into a finite magnetic field on the composite fermions. 

Many aspects of the duality above closely resemble particle vortex duality for bosons.\cite{Dasgupta,Fisher}   Denoting the boson by a complex scalar field $\Phi$, we have the XY action,
\beq L = |(\d_{\mu} - i A_{\mu}) \Phi|^2 + V(|\Phi|) \label{eq:LXY} \eeq
which is dual to the Abelian-Higgs action
\beq L = |(\d_{\mu} - i \alpha_{\mu}) \varphi|^2 + \tilde{V}(|\varphi|) - \frac{1}{2 \pi} \epsilon^{\mu\nu\lambda} A_\mu\partial_\nu a_\lambda \label{eq:AHiggs}\eeq
The dual field $\varphi$  is minimally coupled to a fluctuating electromagnetic field $\alpha$ whose flux is the boson density. Monopole operator of the vortex theory corresponds to $\Phi$, while the dual field $\varphi$ inserts vortices into the bose fluid.  The duality is believed to hold at the critical point of theories (\ref{eq:LXY}), (\ref{eq:AHiggs}), i.e. at the insulator-superfluid transition of bosons. An important question then arises: is the dual surface state in Eq.~(\ref{duaL}) dynamically equivalent to the usual free electron Dirac cone (with chemical potential at the node)? This would be the simplest form of the correspondence, but is not something that we can prove at present. This interesting question is discussed further in Section \ref{sec:dynamics}.

 The existence of the dual surface theory (\ref{duaL}) clarifies a number of earlier mysteries. Two different surface topological orders were put forward for the TI. The first, the Pfaffian-antisemion state consisting of 12 nontrivial anyons, was obtained using a vortex condensation method  \cite{Wang2013a,Metlitski2013}. Another, simpler topological order the T-Pfaffian, with half as many anyons was also proposed \cite{Chen2014PRB,Bonderson2013}, but despite its apparent simplicity could not be `derived' in an analogous fashion, or directly connected back to the superconducting surface state of the topological insulator. We will see that T-Pfaffian is readily derived from the dual surface theory.  Another observation that was previously mysterious  was the close relation between topological superconductors (class AIII) with $\nu=1$ Dirac cone and topological insulators. In both cases there is a U(1) symmetry that can be spontaneously broken at the surface; the statistics of vortices in the resulting surface superfluid can be determined. A striking observation is that vortex statistics on the TI surface is closely related to the STO on the TSc surface, and vice versa. For example, vortices on the TSc surface have the same statistics {  and transformation properties under time-reversal} as the T-Pfaffian topological order. The dual surface theory sheds light on this apparent coincidence. 

We note recent works which have a significant conceptual overlap with the present paper. In Ref. \onlinecite{Mross2015}, Mross, Essin and Alicea explicitly construct a gapless surface state for the TI called the composite Dirac liquid (CDL). Like the present dual Dirac theory, pairing the CDL leads to the T-Pfaffian state. However, in contrast to our dual theory, charge fluctuations are gapped in the CDL, and the gapless Dirac fermions have short ranged interactions. These differences may potentially be bridged with a different choice of interactions for the charge carrying modes. Another insightful development is Son's proposal in Ref. \onlinecite{Son} for a dual description of the particle-hole symmetric half-filled Landau level (see also Ref. \onlinecite{Barkeshli}). At first sight this purely 2d problem seems unrelated to the anomalous surface theories we are discussing in this paper, which always occur on a higher dimensional topological bulk. However, particle-hole symmetry of a Landau level is a nonlocal symmetry. Hence it can  evade restrictions imposed on usual symmetries, and thereby realize the equivalent of an anomalous surface theory in the same dimension. Indeed our work can be viewed as a `derivation' of the conjecture in Ref. \onlinecite{Son}, in a  setting where symmetries are conventionally implemented (such as on the surface of a topological phase where Landau levels with locally implemented particle hole symmetry can be realized). 

{  This paper is organized as follows. We derive the surface QED$_3$ theory in Eqn. \ref{duaL} in two stages. First, in section \ref{sec:bulkpartons}, we present an unconventional construction of a 3d topological insulator. This construction starts with a gapless $u(1)$ spin-liquid phase, with emergent fermionic quasiparticle realizing a topological {\em superconductor} band structure. The 3d TI is then obtained after a confinement transition (see Figure \ref{fig:Derive}). Next, in section \ref{sec:surfpartons} we derive the surface theory that follows from this bulk construction and show it is  given by QED$_3$. 
In section \ref{sec:heur} we present a more heuristic derivation of  QED$_3$ based solely on the surface physics, which provides a transparent physical interpretation of the dual fields. Section \ref{sec:dynamics} discusses possible scenarios for the low energy dynamics of QED$_3$. In section \ref{sec:surfacephases} we show how previously known surface phases of the 3d TI, including the time reversal symmetry broken insulator and Fu-Kane superconductor surface states as well as the surface topological order can be obtained in the dual description. In section \ref{sec:othernu} and  \ref{sec:bulkduality} we show how the particle-vortex duality  of the 2d surface theory can be understood as a descendant of electric-magnetic duality of 3d $u(1)$ gauge theory.}

\section{A parton construction of a topological insulator}
\label{sec:bulkpartons}

In this section we will use parton techniques to construct a 3d gapped state of electrons with no intrinsic topological order and $U(1) \rtimes T$ symmetry. The 2d surface of this state is described exactly by the gauge theory in Eq.~(\ref{duaL}). We will argue that the constructed bulk state is continuously connected to a non-interacting topological insulator, therefore, the theory (\ref{duaL}) provides a description of the TI surface.

The ingredients we will utilize are:
\begin{enumerate}
\item {A trivial ($\theta_{EM} =0$) band insulator of electrons.}
\item  {A spin-liquid state of neutral bosons SL$_\times$.}
\end{enumerate}
While we are ultimately interested in constructing a $T$-invariant state of electrons (fermions) charged under the electromagnetic $U(1)$ symmetry, as a first step we will build a $T$-invariant state of neutral bosons (spins). We will label this state as SL$_{\times}$. This state will be described by an emergent $u(1)$ gauge theory and will possess a gapless photon excitation. One can think of SL$_\times$ as a spin-liquid with global time reversal symmetry (thus the notation SL; we will explain the meaning of the subscript $\times$ shortly). We will then re-introduce the charged electrons and drive a confinement transition in the $u(1)$ gauge theory, obtaining the desired gapped electronic state that is in the same phase as the non-interacting TI.

\subsection{The $u(1)$ spin-liquid of bosons SL$_{\times}$}
\begin{figure}[t!]
\begin{center}
\includegraphics[width = 3.2in]{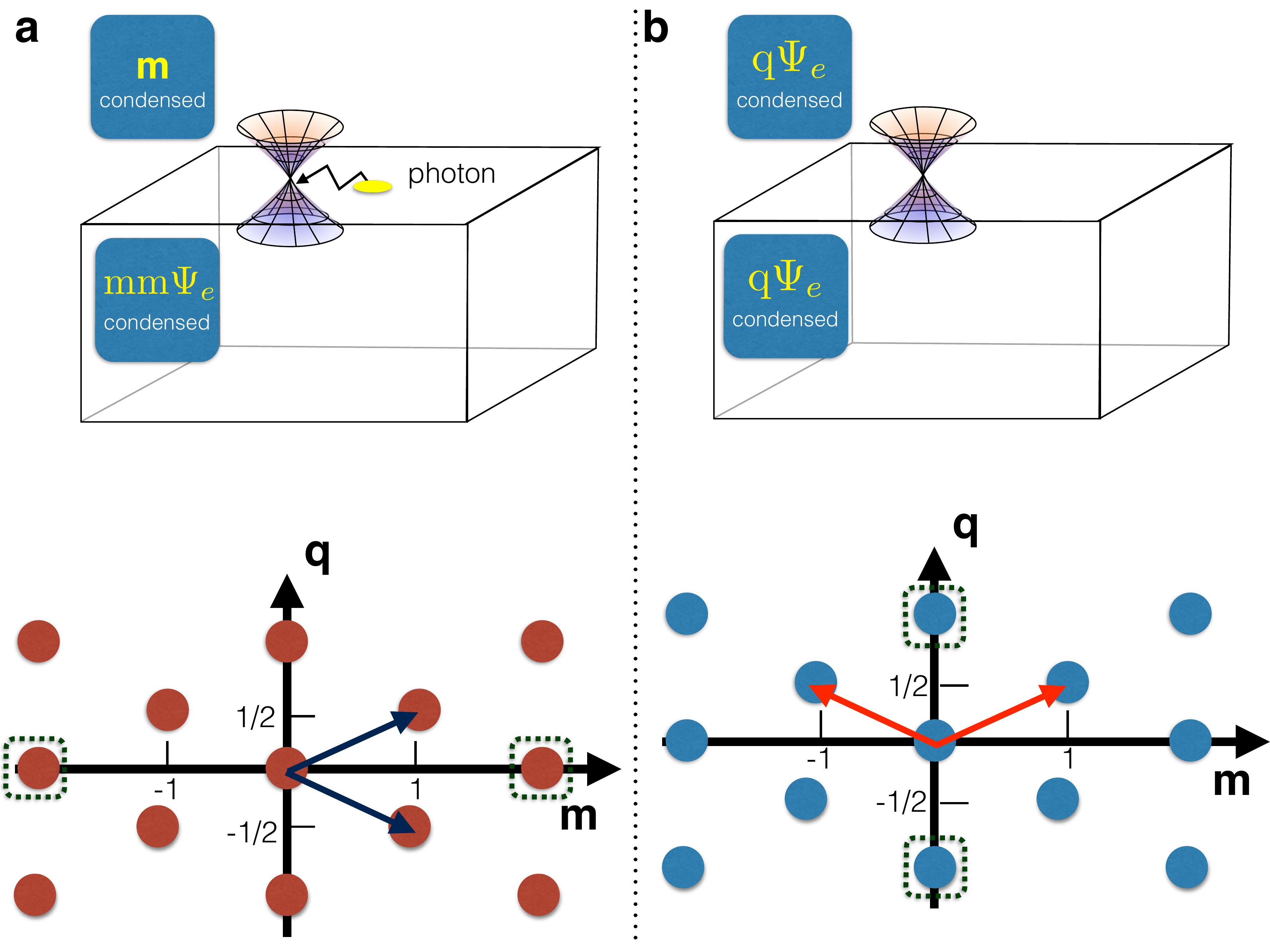}
\vspace{-.2in}
\end{center}
\caption{{\bf (a)} Dual derivation of topological insulator using fermionic partons in a topological {\em superconductor} band structure ($\nu=1$ of class AIII), where time reversal flips the sign of the gauge {\em electric} charge. The bulk topological insulator phase is obtained by condensing a pair of monopoles (0,2) bound to an electron. The surface state consists of the parton Dirac cone coupled to photons that only propagate on the surface, i.e. QED$_3$ {\bf (b)} The 3D TI derived more directly from the partons in the topological insulator band structure, which is Higgsed by condensing the unit electric charge (1,0)  (bound to an electron). The surface is the regular single Dirac cone. The gauged versions of the topological superconductor (a) and topological insulator (b) are related by electric magnetic (E-M) duality - as seen from the lattice of electric and magnetic charges by identifying the basis vectors shown. The two basis vectors are exchanged by time-reversal symmetry in both (a) and (b). The E-M duality relates the double monopole condensate and the Higgs condensate, consistent with obtaining a TI from both descriptions.  }
\label{fig:Derive}
\end{figure}
To construct the spin-liquid state SL$_\times$ we start with a Hilbert space built out of neutral bosons (spins) $B$. We use the standard parton approach where $B$ is decomposed into fermionic constituents $\psi$ as
\beq B = \psi^{\dagger} \Gamma \psi \label{eq:partonB}\eeq
with $\Gamma$ a matrix acting on components of $\psi$. (The precise component structure of $\psi$ and the form of $\Gamma$ will not be important in the discussion below). The representation (\ref{eq:partonB}) is invariant under local $u(1)$ gauge-rotations,
\beq u(1):\,\, \psi(x) \to e^{i \alpha(x)}  \psi(x) \label{eq:partonu}\eeq
The gauge symmetry (\ref{eq:partonu}) will be manifested in the low-energy theory of partons and will give rise to an emergent $u(1)$ gauge field $a_{\mu}$. (We use lower case letters to distinguish the emergent $u(1)$ gauge symmetry from the physical $U(1)$ charge symmetry). 
We take the partons $\psi$ to transform under time-reversal as,
\beq T:\,\, \psi \to U_{T} \psi^{\dagger}, \quad \psi^{\dagger} \to U^*_T \psi \label{eq:Tparton} \eeq
with $U^2_T = -1$, i.e. 
\beq T^2 \psi (T^{\dagger})^2 = - \psi\eeq
Note that the action of $T$ inverts the charge of the partons under the $u(1)$ gauge symmetry. Since $T$ is an anti-unitary symmetry,  $T$ and $u(1)$ commute, so the total symmetry group of the parton theory, which includes both the gauge symmetry and the global $T$ symmetry, is $u(1) \times T$. (Our notation SL$_\times$ emphasizes this direct product structure). Further note that we can combine a rotation by $\alpha$ in the $u(1)$ group with $T$ to get an anti-unitary symmetry $T_{\alpha} = u_{\alpha} T$, which squares to $e^{2 i \alpha}$ when acting on $\psi$. Thus, while we have nominally chosen the partons to transform as Kramers doublets under $T$, this has no physical consequence and is a pure convenience. The fact that the time reversal partners $\psi$ and $\psi^{\dagger}$ have different gauge charge implies that they lie in different topological sectors of the $u(1)$ gauge theory and hence cannot be assigned a physical Kramers parity.

To complete the construction of the SL$_\times$ state, we imagine that the partons $\psi$ form a band-insulator. The transformations of $\psi$ under $u(1) \times T$ are identical to those of an electronic system in class class AIII. One typically thinks of class AIII as $T$-invariant superconductors with a $u(1)$ symmetry, corresponding to the conservation of the $S^z$ components of spin. In our set-up, the $u(1)$ symmetry is an emergent gauge symmetry and has no relation to spin conservation, so the analogy to class AIII superconductors is purely formal. Recall that non-interacting superconductors in class AIII have an integer classification $\nu \in {\mathbb Z}$.\cite{KitaevNI, LudwigNI} The 2d  boundary between the phase with $\nu \neq 0$ and the vacuum ($\nu = 0$) supports $|\nu|$ Dirac cones.  
In the presence of interactions, the non-interacting phases are known to collapse to a $\mathbb{Z}_8$ group.\cite{FidkowskiChenAV, MetlitskiChenFidkowskiAV2014, Wang2014}  Interactions also introduce a single novel phase absent in the non-interacting classification, bringing the total classification in class AIII to ${\mathbb Z}_8 \times {\mathbb Z}_2$.\cite{Wang2014}

To build our spin-liquid state, we will place the partons $\psi$ into a non-interacting band-structure with $\nu = 1$. (We give an example of such a band-structure in appendix \ref{sec:AIII}). Since the partons are gapped, the resulting $u(1)$ gauge theory is in the Coulomb phase with a gapless photon $a_{\mu}$. The effective action of the gauge-field $a_{\mu}$ is given by,
\beq S[a_{\mu}] =  \int d^3 x d t \left( \frac{1}{4 e^2} f_{\mu \nu} f^{\mu \nu} + \frac{\theta}{32 \pi^2}  \epsilon^{\mu \nu \lambda \sigma} f_{\mu \nu} f_{\lambda \sigma}\right) \label{eq:SMaxwella}\eeq
with $\theta=\pi$ and $f_{\mu \nu} = \d_{\mu} a_{\nu} - \d_{\nu} a_{\mu}$. The first term in Eq.~(\ref{eq:SMaxwella}) is the Maxwell action with a coupling constant $e$. The second term is the topological term generated by integrating out the partons. Similarly to electrons in an ordinary TI, for partons in a $\nu = 1$ band-structure the coefficient $\theta = \pi$. 

Let us discuss the spectrum of topological excitations in the SL$_{\times}$ state. One type of topological excitations is given by partons $\psi$ - i.e. electric charges of $a_{\mu}$. The second type of excitations is given by magnetic monopoles of $a_{\mu}$. The presence of the topological term (\ref{eq:SMaxwella}) endows a monopole with magnetic flux $2 \pi m$ with an electric charge\cite{WittenEffect} 
\beq q = n + \frac{\theta m}{2 \pi} = n + \frac{m}{2} \label{eq:WittenEffect}\eeq
 Here $n$ is an arbitrary integer which reflects the freedom of adding $n$ electric charges $\psi$ to a monopole. Thus, the topologically distinct excitations form a 2-dimensional lattice $(q, m)$ labelled by electric charge $q$ and magnetic charge $m$ satisfying Eq.~(\ref{eq:WittenEffect}), see Fig.~\ref{fig:Derive}a.

The excitations have the following statistics. A single charge $\psi = (1, 0)$ is a fermion. A single monopole $(q, m = 1)$ of arbitrary electric charge $q$ is a boson. A general dyon $(q, m)$ has statistics $(-1)^{(q-m/2) (m+1)}$ with $+1$ corresponding to bosonic statistics and $-1$ to fermionic. In particular, a neutral double monopole $(q = 0, m = 2)$ is a fermion.  Two dyons $(q, m)$ and $(q', m')$ experience a non-trivial statistical interaction. Namely, if we place $(q', m')$ at the origin and let $(q, m)$ move along a closed contour ${\cal C}$, $(q, m)$ will accumulate a statistical phase $(q m' - q' m) \Omega/2$, where $\Omega$ is the solid angle subtended by ${\cal C}$. 

In addition to the statistical interaction, dyons also experience a $1/r$ Coulomb interaction. 
As already mentioned, time-reversal symmetry (\ref{eq:Tparton}) maps electric charge $q \to -q$. Furthermore, due to the $u(1) \times T$ group structure the magnetic flux is preserved under $T$: $m \to m$. Let us next discuss the physical Kramers parity $T^2$ of the excitations. Kramers parity can only be assigned to topological sectors which are left invariant under $T$. In the present case, this corresponds to $(q = 0, m)$ with $m$-even. As has been discussed in Refs.~\onlinecite{MetlitskiChenFidkowskiAV2014, Wang2014},  the neutral double monopole $(q = 0, m = 2)$ is actually a Kramers doublet fermion. This is required by the consistency of the theory since $(0, 2)$ can be obtained by fusing the time-reversal partners $(1/2, 1)$ and $(-1/2, 1)$. The presence of a non-trivial statistical interaction between $(1/2, 1)$ and $(-1/2, 1)$ forces their fusion product $(0, 2)$ to be a Kramers doublet fermion.\cite{Metlitski2013, ChongScience}

Next, let us discuss the surface of SL$_{\times}$ phase. Let us imagine that space is divided into two regions by an interface at $z = 0$. We will place our partons $\psi$ into the $\nu = 1$ band-structure for $z < 0$ and into the trivial $\nu = 0$ band-structure for $z > 0$. The interface then supports a single gapless Dirac cone of $\psi$,
\beq S_{2d} = \int d^2 x d t \, \bar{\psi}_{cf} i \gamma^{\mu} (\d_{\mu} - i a_{\mu}) \psi_{cf} \label{eq:Diracgauged} \eeq
with $\psi_{cf}$ now denoting the surface Dirac fermion. Under $T$, $\psi_{cf}$ transforms as
\beq T:\,\, \psi_{cf} \to \epsilon \psi^{\dagger}_{cf}\eeq
where $\epsilon = i \sigma^y$ and we are using the basis of $\gamma$ matrices $(\gamma^0,\, \gamma^1,\,\gamma^2) = (\sigma_y,\, -i\sigma_z,\, i\sigma_x)$. Again, we stress that this is different from the $T$-transformation of the electron on the free Dirac surface of a TI (\ref{eq:LDirace}),
\beq T:\,\, \Psi_e \to \epsilon \Psi_e\eeq

The $z < 0$ region realizes our SL$_{\times}$ phase. The $z > 0$ also realizes a spin-liquid described by a $u(1)$ gauge-theory with a gapless photon. Let us briefly discuss the properties of the spin liquid in the $z > 0$ region.  Since here the partons are in a trivial band-structure, the topological angle $\theta = 0$, and the dyon spectrum is given by $(q, m)$ with $q$ - integer and $m$ - integer. The $(q, m)$ dyon  has statistics $(-1)^{q (m+1)}$, in particular, all neutral monopoles are bosons.  Time reversal symmetry again acts as  $T: \, (q, m) \to (-q, m)$, however,  the single monopole $(0, 1)$ is now a Kramers singlet. 

So far, we have constructed an interface between two $u(1)$ spin-liquid phases: one with $\theta = 0$ and one with $\theta = \pi$. The 2d  gapless Dirac fermion appearing on the interface (\ref{eq:Diracgauged}) interacts with a 3d gapless photon living on both sides of the interface. In order to construct an interface between the spin-liquid with $\theta = \pi$ and the vacuum, we need to drive a confinement transition in the region $z > 0$. This can be done by condensing the single neutral monopole $(0, 1)$ in the region $z > 0$. Since this monopole is a boson it can condense. Furthermore, since the monopole is a Kramers singlet the condensation process preserves the $T$-symmetry. 
The only deconfined excitations are $(0, m)$, and since these are multiples of the condensed monopole $(0,1)$ the resulting phase has no topologically non-trivial excitations. Hence, after monopole condensation the $z > 0$ region realizes the trivial $T$-invariant vacuum phase. The Dirac cone on the interface survives the monopole condensation, however, it now interacts with a gauge field $a_{\mu}$, which lives only in the $z < 0$ region.

\subsection{Confinement to a topological insulator}

We next describe how to confine the SL$_\times$ spin-liquid phase with $\theta = \pi$ described in the previous section to a $T$-symmetric, fully gapped insulator of electrons with no intrinsic topological order. As a first step, we will now need to work in a Hilbert space which includes the physical charged electron $\Psi_e$ which is a Kramers doublet under time reversal symmetry. 


Let us begin by taking a non-interacting ``mixture" of a trivial band insulator of electrons and the SL$_\times$ state of neutral bosons with $\theta = \pi$ constructed in the previous section. Consider a bound state $D$ of the electron $\Psi_e$ and the neutral double monopole of the spin-liquid $(q = 0, m = 2)$. As we discussed above, $(0, 2)$ is a Kramers doublet fermion. Therefore, $D$ is a Kramers singlet boson, which can condense preserving $T$. What are the properties of the resulting phase? Recall that generally condensation of a dyon with charges $(q, m)$ gives rise to an analogue of a Meissner effect for the gauge field combination $q \vec{b} - 2 \pi m \vec{e}$, with $\vec{b} = \nabla \times \vec{a}$ - the magnetic field of $a_{\mu}$, and $\vec{e} =  \d_t \vec{a} - \nabla a_t$ - the electric field of $a_{\mu}$. All excitations which are sources of this gauge field combination will be confined, i.e. a dyon $(q', m')$ is confined if $q m' - m q' \neq 0$ (i.e. only dyons which possess trivial mutual statistics with $(q,m)$ are deconfined). Now, since the electron has no charge under $a_{\mu}$, our condensing dyon $D$ still has electric gauge charge $q = 0$ and magnetic charge $m = 2$. Its condensation will gap out the photon giving rise to the ``Meissner" effect $\vec{e} = 0$. Therefore, all excitation with gauge charge $q \neq 0$ will be confined. Remembering that $q = n + m/2$,  only excitations with $q = 0$ and  $m$ - even are deconfined. These excitations are multiples of the condensing dyon $D$ (possibly with electrons $\Psi_e$ added on top). Therefore, the condensed phase has no non-trivial deconfined excitations and so possesses no intrinsic topological order.  

What is the fate of the physical $U(1)$ charge symmetry in the $D$-condensed phase? First, all excitations in the Coulomb phase can be labelled by $(q, m; Q)$, with $(q, m)$ being the emergent $u(1)$ electric and magnetic quantum numbers coming from the SL$_\times$ sector, and $Q$ being the physical $U(1)$ charge coming from the electron band-insulator sector. 
Nominally, $D$ has quantum numbers $(q = 0, m = 2; Q = 1)$. Therefore, one might naively think that the $D$-condensed phase breaks the $U(1)$ symmetry and is a superfluid. However, this is not the case. Indeed, one cannot build any local observable (i.e. one with $q = 0$ and $m = 0$) with non-zero $Q$ out of $D$. {  More physically, recall that the dyons $D$ experience a long-range $1/r$ interaction.} 
In the $D$-condensed phase, the dyons $D$ form a Debye-plasma with short-range correlations, so the resulting state is gapped. This fact is completely insensitive to $D$'s carrying a global $U(1)$ charge.  So the $D$-condensed phase cannot possibly be a superfluid, since a superfluid would necessarily possess a gapless Goldstone mode; rather, it is an insulator. Now, let us imagine inserting a gapped double monopole $(0, 2)$ with $Q = 0$ into the Debye plasma of $D$'s. This double monopole will be Debye screened by the $D$'s - it will be surrounded by a cloud of $D$'s and $D^{\dagger}$'s with a total $D$-number equal to $-1$.  Now, as each $D$ carries a $U(1)$ charge $Q = 1$, the screening cloud has a total electric charge $Q = -1$. Therefore, we conclude that a deconfined double monopole sucks up a physical electric charge $-1$ in the $D$-condensed phase. 
More generally, the true physical $U(1)$ charge of an excitation with ``nominal" quantum numbers $(q, m; Q)$ in the $D$-condensed phase is,
\beq Q_{EM} = Q - m/2 \label{eq:QEM}\eeq
Note that since only dyons with even $m$ are deconfined, the electric charge $Q_{EM}$ is always an integer.

Let us now argue that this phase has a response to the {  physical electromagnetic field $A_{\mu}$} characterized by $\theta_{EM}=\pi$. This is most conveniently done via the Witten effect, by calculating $Q_{EM}$ of an inserted monopole of $A_{\mu}$. 
Before the $D$-condensation, we can label all excitations by $(q, m; Q, M)$ where $M$ now represents the magnetic charge under $A_{\mu}$. Since the response of our initial Coulomb phase to $A_{\mu}$ comes entirely from the trivial electron band insulator, the $U(1)$ sector is characterized by a $\theta$ angle $\theta_{EM} = 0$, so  $Q$ and $M$ are both integers. Now, $D$ has quantum numbers $(q = 0, m = 2; Q = 1, M = 0)$, so its condensation leads to a Meissner effect for $4 \pi \vec{e} -  \vec{B}$, where $\vec{B} = \nabla \times \vec{A}$ is the magnetic field strength of $A_{\mu}$. Thus, deconfined excitations must have $2 q = M$. In particular, an $M = 1$ external monopole of $A_{\mu}$ must carry $a_{\mu}$ electric charge $q = 1/2$. Since $q - m/2$ is an integer, we conclude that the $M = 1$ external monopole must bind an odd number $m$ of monopoles of $a_{\mu}$. 
{  From Eq.~(\ref{eq:QEM}),} we  conclude that an external $M = 1$ monopole will bind a half-odd-integer physical electric charge $Q_{EM}$. This implies that the $D$-condensed phase has a response to the external $U(1)$ gauge field with $\theta_{EM} = \pi$ - the same as the EM response of a non-interacting topological insulator.

By an argument of Ref.~\onlinecite{ChongScience}, a phase of electrons with $\theta_{EM} = \pi$ is identical to a non-interacting topological insulator up to an SPT phase of neutral bosons with $T$-invariance. 
In section \ref{sec:bulkduality}, we will arrive at the same conclusion without using the general argument of Ref.~\onlinecite{ChongScience}. Moreover, by strengthening the argument in section \ref{sec:bulkduality}, we will be able to show that the $D$-condensed phase is continuously connected to a non-interacting TI with no additional bosonic SPT.\cite{Metlitski_ta}

\section{The surface theory - QED$_3$}

\subsection{Derivation of surface theory from parton construction}
\label{sec:surfpartons}
Let us now turn to the surface of the $D$-condensed phase. As before, we imagine putting partons into a $\nu = 1$ band-structure of class AIII in the region $z < 0$ and into the trivial $\nu = 0$ band-structure in the region $z > 0$.  The electrons $\Psi_e$ are placed into a trivial band-insulator band-structure everywhere in space. The interface at $z =0$ supports a single Dirac cone of partons $\psi_{cf}$ coupled to a bulk $u(1)$ gauge-field $a_{\mu}$ living on both sides of the interface, see Eq.~(\ref{eq:Diracgauged}). We drive the $z > 0$ side of the interface into a trivial insulating state (vacuum) by condensing the single monopole $(q = 0, m = 1; Q = 0)$. We condense the $D$-dyon on the $z < 0$ side of the interface, driving it into the topological insulator phase. Both sides of the interface are now confined and exhibit the Meissner effect, $\vec{e} = 0$. The Dirac cone of partons $\psi_{cf}$ on the surface survives the condensation, since  the bulk gap to partons $\psi$ persists during the condensation process. Due to the bulk Meissner effect, the field lines of the electric field $\vec{e}$ cannot penetrate into the bulk and can only stretch along the surface.  On the other hand, a surface magnetic field $b_z$ perpendicular to the interface is allowed - such a magnetic field gets Debye screened by the condensed monopoles/dyons on both sides of the interface. Thus, the gauge field $a_{\mu}$ is now confined to live on the surface becoming a  2+1 dimensional $u(1)$ gauge-field, so the surface theory is simply given by QED$_3$ with a single flavor of Dirac fermions.

Let us discuss the response of the surface to the $U(1)$ gauge field $A_{\mu}$. Imagine that there is a magnetic field $b_z = \d_x a_y - \d_y a_x$ perpendicular to the surface. As already noted, this magnetic field will be Debye screened by monopoles/dyons on the two sides of the interface. On the $z > 0$ side, the condensed monopoles $(q = 0, m = 1; Q = 0)$ will form a screening layer with 2d density $\rho_m = -\frac{1}{2\pi} b_z$. Since these monopoles carry no $U(1)$ charge, they do not contribute to the physical electric charge density. On the $z < 0$ side, the condensed $D$-dyons (quantum numbers $(q = 0, m = 2; Q = 1)$) will form a screening layer with 2d density $\rho_{D} = \frac{1}{4 \pi} b_z$. Since each $D$ has electric charge $Q = 1$, this screening layer creates a $U(1)$ charge density $\rho_{EM} = \frac{1}{4\pi} b_z$. 

Similarly, imagine that an electric field $e_i$ ($i = x, y$) along the interface is present. This electric field must be Meissner screened by monopole currents on both sides of the interface (analagous to how a magnetic field along the surface of a superconductor is Meissner screened by electric currents). On the $z > 0$ side of the interface this results in a monopole current $j^m_i = \frac{1}{2 \pi} \epsilon_{ij} e_j$. Since these monopoles are neutral, the monopole current does not contribute to the $U(1)$ current. On the $z < 0$ side of the interface the electric field is screened by a current of $D$-dyons, $j^D_i = -\frac{1}{4 \pi} \epsilon_{ij} e_j$, which translates into a $U(1)$ electric current $j^{EM}_i = -\frac{1}{4 \pi} \epsilon_{ij} e_j$. Thus, we conclude that the surface gauge-field $a_{\mu}$ induces a $U(1)$ current $j^{\mu}_{EM} = \frac{1}{4 \pi} \epsilon^{\mu \nu \lambda} \d_{\nu} a_{\lambda}$, and the effective action of the surface theory in the presence of an external $U(1)$ gauge field $A_{\mu}$ is
\beq L_{QED_3} =  \bar{\psi}_{cf} i \gamma^{\mu} (\d_{\mu} - i a_{\mu}) \psi_{cf}  - \frac{1}{2 (2\pi)} \epsilon^{\mu \nu \lambda} A_{\mu} \d_{\nu} a_{\lambda} \label{eq:DiracgaugedA} \eeq


We immediately see that dynamical instantons of $a_{\mu}$ are prohibited in the surface theory as they do not preserve the $U(1)$ charge. However, instantons of flux $\phi = 2 \pi m$, with $m$ - even,  do correspond to physical operators with electric charge $Q_{EM} = m/2$ in the surface theory. In fact, a flux $4 \pi$ instanton is identified with the electron insertion operator $\Psi_e$. To see this, imagine we create an electron $\Psi_e$ on the surface. Our parton construction had $\Psi_e$ gapped everywhere (including on the boundary). However, $\Psi_e$ can decay into gapless boundary degrees of freedom as follows: it can grab a double monopole from the $z > 0$ side of the interface (where monopoles are condensed) and tunnel across the interface to the $z < 0$ region vanishing into the condensate of $D = \Psi_e \times (0,2)$. An $a_{\mu}$ flux of $4\pi$ and $U(1)$ charge $Q_{EM} = 1$ is created on the surface in the process. Thus, an electron creation operator $\Psi_e$ corresponds to a flux $4 \pi$ instanton in the surface theory.

Note that single (flux $2\pi$) instanton events do not correspond to physical operators in the surface theory. Indeed, a flux $2 \pi$ instanton would correspond to a single monopole tunneling across the $z = 0$ interface. But single monopoles are confined in the $D$-condensed region, so single instanton events are not local operators in the surface theory.

A complementary picture to the above discussion can be obtained by studying instanton events directly in surface QED$_3$ theory. Let us imagine that the TI phase obtained by $D$-condensation occupies a solid ball of radius $R$ and the trivial vacuum occupies the region outside this ball. The surface theory then lives on a sphere $S^2$. A strength $m$ instanton event in the surface theory will create a flux $2 \pi m$ on the surface. For simplicity, imagine this flux spreads uniformly across the surface. The single-particle spectrum of a Dirac fermion on $S^2$ in the background of a uniform flux $2 \pi m$ possesses $N_0 = m$ zero modes. Recall that time-reversal symmetry inverts the $u(1)$ charge density, $T \rho(x) T^{\dagger} = - \rho(x)$. This implies that the total $u(1)$ charge of a state with all the negative energy modes filled and all the zero and positive energy modes empty is $q = -N_0/2 = -m/2$.\cite{KapustinMonopoles} On a compact space such as $S^2$ the total $u(1)$ gauge charge $q$ must be zero. Therefore, we must fill $m/2$ out of $m$ zero energy modes. If $m = 1$, there is only a single zero-energy level, so we cannot half-fill it: the state with the zero mode empty has $q = -1/2$ and the state with the zero mode filled has $q = 1/2$. Therefore, the flux $2 \pi$ instanton is not a local operator in QED$_3$. However, if $m = 2$, we have two zero modes: we can fill either one or the other, obtaining two degenerate $q = 0$ states. In fact, these states transform in the $j = 1/2$ representation of the rotation group on the sphere. Since the surface theory QED$_3$ is Lorentz invariant, the fact that we've changed the spin of the system by $1/2$ implies that we have added a fermion to the system. This is consistent with our identification of the flux $4\pi$ instanton operator with the electron $\Psi_e$.

It is often stated  that a single Dirac fermion in $2+1$ dimensions suffers from the parity anomaly, namely it cannot be consistently coupled to a $u(1)$ gauge field preserving the time-reversal symmetry.\cite{Redlich} Our surface theory (\ref{eq:DiracgaugedA}) has a dynamical $u(1)$ gauge field $a_{\mu}$, which is confined to live just on the surface (we switch off the background electromagnetic field $A_{\mu}$ for now). Thus, the standard argument for the evasion of the anomaly via the $3+1$ dimensional bulk does not directly apply in this case. Rather, the anomaly is resolved by modifying the compactification of the gauge-field $a_{\mu}$ in the surface theory. This important point is discussed in detail in Appendix \ref{App:Compactness}. 

\subsection{Heuristic derivation of dual surface theory}
\label{sec:heur}
We now give a more heuristic derivation of the dual surface QED$_3$ theory, which starts directly with the Dirac cone of electrons (\ref{eq:LDirace}) and does not rely on the bulk construction in Section \ref{sec:bulkpartons}. 

Consider the TI surface state (single Dirac cone) with U(1) charge and time reversal symmetry U(1)$\rtimes T$. For simplicity we assume that the chemical potential is at the Dirac point. Now, consider surface superconductivity and the statistics of vortices induced in the superconductor. This was studied in Refs.~\onlinecite{Metlitski2013, Wang2013a} where the  vortex theory in Table \ref{vortex} was derived. The vortex statistics can be described by a TQFT Ising$\times U(1)_{-8}$. The anyons  in this TQFT are labelled by the Ising charge $\{I, \sigma, \psi\}$ and a $U(1)_{-8}$ charge $k$, which will be noted as a subscript on the Ising charge. The $U(1)_{-8}$ charge coincides with the vorticity (the $hc/2e$ vortex is the unit vortex). Not all sectors of Ising$\times U(1)_{-8}$ TQFT are realized by vortices: vorticies with odd vorticity always have an Ising charge $\sigma$, and vorticies with even vorticity have an Ising charge $I$ or $\psi$.
Time-reversal symmetry reverses the vorticity. 

Note that the Ising$\times U(1)_{-8}$ TQFT is the same as one describing the T-Pfaffian surface topological order of a TI (table \ref{TPfaffian}), however, the action of time-reversal symmetry on the $U(1)_{-8}$ charge $k$ is different. In a T-Pfaffian state  $k$ is preserved by $T$, while in the vortex theory it is reversed. A further difference is that anyons in the T-Pfaffian state also carry charge $k/4$ under the physical $U(1)_{EM}$ global symmetry. Let us define the CT-Pfaffian to be a topological order of electrons with $U(1) \times T$ global symmetry with the same anyon content and $U(1)$ charge assignments as the T-Pfaffian, but reversed action of time-reversal symmetry on the $U(1)_{-8}$ charge. One may suspect that the CT-Pfaffian is the surface topological order of the $\nu = 1$ class AIII topological superconductor. Indeed, precisely this proposal has been made in  Ref.~\onlinecite{FidkowskiChenAV} (modulo a bosonic SPT phase, which we will return to). Note that the fermion $\psi_4$ in the $T$-Pfaffian and CT-Pfaffian states has trivial mutual statistics with all other anyons and is identified with the physical electron.

\begin{table}[t]
\beq
\begin{array}{|c|c|c|c|c|c|c|c|c|}
\hline
k\rightarrow     & 0 & 1 & 2 & 3 & 4 & 5 & 6 & 7 \\ \hline
I    & 1&   & \,-i&   &1\ &   &\,-i&   \\ \hline
\sigma  &   & \,\,1\,\, &   & -1 &   &  -1 &   &  \,\,1\,\,  \\ \hline
\psi & -1 &   & i&   & {\color{red} -1} &   & i &  \\ \hline 
\end{array}\nn\eeq
\caption{Vortex defects on the surface of a topological insulator. The vortex statistics are described by Ising$\times U(1)_{-8}$ theory and the table lists the topological spins of vortices. The column index is the flux $k\frac{h c}{2e}$, which coincides with the $U(1)_{-8}$ charge and the row index is the Ising charge. 
Only $\psi_0$ has a well-defined Kramers parity $T^2 = -1$. The fermionic vortex $\psi_4$ has trivial mutual statistics with all other vortices. }
\label{vortex}
\end{table}

\begin{table}[t]

\beq
\begin{array}{|c|c|c|c|c|c|c|c|c|}
\hline
k\rightarrow     & 0 & 1 & 2 & 3 & 4 & 5 & 6 & 7 \\ \hline
I    & 1&   & \,-i&   &1\ &   &\,-i&   \\ \hline
\sigma  &   & \,\,1\,\, &   & -1 &   &  -1 &   &  \,\,1\,\,  \\ \hline
\psi & -1 &   & i&   & -1 &   & i &  \\ \hline\hline 
T^2 & 1& \eta & & -\eta & -1 & -\eta& & \eta\\ \hline
\end{array}\nn\eeq
\caption{T-Pfaffian$_\eta$ topological orders with $\eta = \pm 1$. The top table lists the topological spins of anyons; the column and row indices denote the $U(1)_{-8}$ charge and the Ising charge respectively. The physical electric charge of anyons $Q_{EM} = k/4$ with $k$ - the $U(1)_{-8}$ charge. Time-reversal maps $k$ to itself.  The bottom row lists the $T^2$ assignment of anyons (where defined). The $T^2$ assignment is independent of the Ising charge.  $\psi_4$ is the physical electron. The CT-Pfaffian topological order has identical anyon content and charge assignments, but $T$ maps $k \to -k$ and $\psi_0$ has $T^2 = -1$. }
\label{TPfaffian}

\end{table}

Let us come back to the vortices on the surface of a TI. To encode logarithmic interactions between vorticies we can take them to carry gauge charge of an emergent $u(1)$ gauge field $a_{\mu}$ (in addition to charges in the Ising$\times U(1)_{-8}$ TQFT). For notational convenience, let us normalize the charge of a unit vortex under $a_{\mu}$ to be $1/4$. Let us for a moment ignore the fluctuations of $a_{\mu}$, and treat the vorticity as a charge under a global $u(1)$ symmetry. 
Then we can regard the vortex theory as a topological order CT-Pfaffian with $u(1) \times T$ symmetry, which is the surface state of the $\nu = 1$ TSc in class AIII. In this identification, the quadruple vortex $\psi_4$ is identified with the electron of the class AIII TSc.  

With this connection in hand, we can discuss other possible surface states of the class AIII TSc.  The simplest one of course is the single Dirac cone, which will now be composed of the $\psi_4$ fermions. Reinstating the fact that the $u(1)$ charge is actually gauge charge, implies that the fermions are coupled to a gauge field as described by  the Lagrangian   (\ref{duaL}). This line of reasoning provides a physical picture of the dual fermionic field $\psi_{cf} = \psi_4$, i.e. it is the strength 4 fermionic vortex on the surface of a topological insulator. 
Also, if one prefers to view the vortex as a boson, then this is a bound state of a 4$\frac{hc}{2e}$ vortex and a neutral Bogoliubov quasiparticle.   Due to its vorticity,  it is minimally coupled to a gauge field $a$. Since each electron appears as a flux of 4$\pi$ to a strength 4 superconductor vortex ($2hc/e$ vortex), we have $(\partial_x a_y-\partial_y a_x)  = 4\pi\rho_e$ - the electron density. The terminology `composite fermion' to describe the fermionic vortex $\psi_4$ should be clear now.  Recall, in the original definition of composite fermions each electron is attached to a pair of  $hc/e$ vortices - as in $\psi_4$. While this is usually discussed in the context of quantum Hall states\cite{HLR,Jain}, the present discussion shows that this duality is also relevant to describe the surface of a topological insulator.

The Dirac theory of composite fermions apparently represents a different gapless surface state from the original Dirac model - however, the simplest conclusion that the two models are dynamically equivalent, also remains an intriguing possibility as discussed below. 

\subsection{Dynamics of QED$_3$}
\label{sec:dynamics}
So far we have ignored the issue of the dynamics of the surface QED$_3$ theory. Instead, our discussion was focused on issues of symmetries and quantum numbers of operators. In principle, the surface theory (\ref{eq:DiracgaugedA}) can be perturbed by an orbitrary local symmetry preserving operator. This gives rise to a large landscape of possible surface phases, some of which will be discussed in section \ref{sec:surfacephases}. Since the $D$-condensed phase in our bulk construction is continuously connected to a non-interacting TI, one of these surface phases must be the gapless (ungauged) free Dirac cone of electrons. Thus, the Dirac cone is dual to QED$_3$ in the sense of duality of Hilbert spaces, operators and symmetries. 

One can ask whether a stronger version of duality holds. Namely, is weakly coupled QED$_3$ dual in the infra-red to the free Dirac cone. By  weakly coupled QED$_3$ we mean the theory
\beq L_{QED_3} =  \bar{\psi}_{cf} i \gamma^{\mu} (\d_{\mu} - i a_{\mu}) \psi_{cf}  + \frac{1}{4 g^2} f_{\mu \nu} f^{\mu \nu}  \label{eq:QED3UV}\eeq
with the coupling constant $g^2$ much smaller than the UV cut-off $\Lambda_{UV}$. The theory (\ref{eq:QED3UV}) is well defined and analytically controlled in the UV (i.e. for energy scales $g^2 \ll \omega \ll \Lambda_{UV}$) where it reduces to a Dirac fermion $\psi_{cf}$ interacting weakly with a massless photon $a_{\mu}$. The fate of the theory in the IR is not known. One can envision three different scenarios, which we list here in the order of increasing exoticity:

i) The theory (\ref{eq:QED3UV}) spontaneously breaks time-reversal symmetry in the infra-red, dynamically generating a fermion mass term $m \bar{\psi}_{cf} \psi_{cf}$. The IR theory is then a trivial gapped state with no intrinsic topological order {  (see section \ref{sec:Tbreak} and Appendix \ref{App:Compactness})}. It is identical to the phase obtained from the free Dirac cone by spontaneously breaking $T$-symmetry and generating a mass $m \bar{\Psi}_e \Psi_e$. This is the most conventional scenario. In fact, the standard (although unproven) expectation is that non-compact QED$_3$ with a small number of fermion flavors $N_f$ and a symmetry group $SU(N_f)$ does generate a fermion mass in the IR.  However, as we review in appendix \ref{App:Compactness}, conventional 2+1 dimensional QED$_3$ with the standard large gauge transformations is only consistent with $T$-symmetry for even $N_f$. For odd $N_f$, one must add a Chern-Simons term with a half-odd-integer level $k$ to the massless Dirac theory in order to preserve the standard large gauge transformations (see appendix \ref{App:Compactness}). In this case, the $T$-symmetry is absent so a fermion mass term is allowed by symmetry and is, in fact, generated already in perturbation theory. Now, our $N_f  =1$ surface theory has no Chern-Simons term due to the modification of allowed large gauge transformations. Therefore, it does not fit into the conventional folklore and perhaps can remain gapless in the IR. This brings us to the other two scenarios.

ii) Weakly coupled QED$_3$ with $N_f = 1$ flows in the infra-red to a free Dirac theory with $N_f = 1$. This would be a ``strong" version of particle-vortex duality for fermions. Such a strong version of duality does, indeed, hold for bosons.

iii) Weakly coupled QED$_3$ with $N_f = 1$ flows in the infra-red to a CFT distinct from a free Dirac theory. 

At present, we can make no statement regarding which of these scenarios is realized.

\section{Dual Descriptions of the TI Surface phases}
\label{sec:surfacephases}
Let us describe how the different phases of the topological insulator surface are realized in the dual fermion description. 

Let us begin by considering symmetry preserving surface states, and then discuss those that break symmetry. This discussion has significant overlap with a previous paper by Son \cite{Son}, with some relabeling of time reversal and particle hole symmetry, and the physical context.  Also Ref. \onlinecite{Son} adopts a minimal coupling of dual fermions to gauge field with twice the charge used in this paper. 

\subsection{Surface phases preserving all symmetries} 

\subsubsection{Fermi liquid and HLR state of dual fermions} 

Consider the situation when we preserve physical symmetries, charge conservation and time reversal, with the metallic surface state of the topological insulator. Typically the metallic surface state will be at finite filling, implying the chemical potential is away from the Dirac node. How is this Fermi liquid state described in the dual language? The finite chemical potential on the electrons implies a finite magnetic field on the composite fermions via the equation $(\partial_x a_y-\partial_y a_x)  = 4\pi\rho_e$. Since the dual  fermions are at neutrality, a consequence of physical time reversal symmetry, they will be at half filling of the zeroth Landau level. Fermions in a magnetic field which fill half a Landau level can be in a variety of states.  One possibility is the `composite Fermi liquid' or  Halperin-Lee-Read \cite{HLR} state. Here, however, since the fermions are themselves composite fermions, performing the duality twice leads us back to the original electrons. This is nothing but the original Fermi liquid of electrons. 


\subsubsection{Surface topological order and superconductivity of dual fermions} 

Another possible way to preserve charge $U(1)$ and $T$ on the TI surface is through surface topological order. Two different symmetric topologically ordered surface states of a TI have been identified. The first is the Pfaffian-antisemion state, with 12 particles, and the other is the T-Pfaffian state, with 6 particles (see table \ref{TPfaffian}). 
While the Pfaffian-antisemion state can be derived by a vortex condensation argument starting from the single Dirac cone surface state of a topological insulator, the T-Pfaffian cannot be analogously derived (see, however, Ref.~\onlinecite{Mross2015}). Rather it is argued to be a consistent surface termination that captures all relevant anomalies of the TI surface. Here, we are able to derive this surface topological order directly from the duality (\ref{duaL}). {  To do so}, we simply consider the composite fermions to be in the Higgs phase, where they have paired and condensed. This gaps out the photon $a_{\mu}$, and the vortices in this phase trap quantized gauge flux $\int d^2x \left (  \partial_x a_y-\partial_y a_x \right )= \pi$. 
This simply means that the unit vortex carries electric charge $Q_{EM} = 1/4$ (since from (\ref{duaL}) a flux $2 \pi$ of $a$ corresponds to charge $1/2$). The statistics of vortices in this superfluid have previously been worked out \cite{MetlitskiChenFidkowskiAV2014,Wang2014} -  and they precisely corresponds to  the T-Pfaffian state, with the same transformation properties under time reversal as proposed in \cite{Chen2014PRB,Bonderson2013}! In fact, the T-Pfaffian appears in two varieties, T-Pfaffian$_\pm$ which differ in the transformation properties under time reversal (see table \ref{TPfaffian}). While one of them corresponds to the topological insulator surface, the other differs from it by addition of the eTmT SPT state of neutral bosons. The eTmT phase admits a toric code surface state where both the $e$ and $m$ anyons are Kramers doublets. Previously, the exact correspondence was unknown. Now, pair condensation of composite fermions in the dual theory (\ref{duaL}), in fact, gives rise to the T-Pfaffian$_+$ topological order.\cite{MetlitskiChenFidkowskiAV2014} Thus, the duality allows to resolve the long-standing T-Pfaffian$_+$/T-Pfaffian$_-$ puzzle. We, however, remind the reader that to establish the duality (\ref{duaL}) we had to argue that the bulk phase constructed in section \ref{sec:bulkpartons} is continously connected to the non-interacting TI and does not differ from it by an eTmT state. The details of this argument will be given in section \ref{sec:bulkduality} and in a forthcoming publication.\cite{Metlitski_ta} 



\subsection{Breaking symmetries}
Now we consider surface phases that break symmetry. 

\subsubsection{Surface superfluid and dual surface topological order}
When electrons pair and condense to form a surface superfluid, we have noted that surface vortices have the same statistics and transformation properties under $T$ as the CT-Pfaffian topological order.
As discussed in section \ref{sec:heur}, this can simply be interpreted as the topological order of composite fermions $\psi_{cf}$ of the dual surface theory. 
An additional feature here is the coupling to the gauge field. When all gauge charges are gapped the photon is free to propagate, which is just the dual description of the Goldstone mode of the electronic superfluid.

A different way to motivate this connection is the following. Consider the metallic surface state of electrons at finite chemical potential. A natural instability of this Fermi liquid is the BCS instability towards pairing. In the dual description, this corresponds to a finite effective magnetic field applied to a half filled Landau level of $\psi_{cf}$. A natural consequence are various non-Abelian topological orders such as Pfaffian, anti-Pfaffian  etc. These, however,  break   particle-hole symmetry, which is just the time reversal symmetry of the electrons. A topological order that preserves particle-hole symmetry is the  CT-Pfaffian, which corresponds to the surface superfluid of electrons.

\subsubsection{Breaking time-reversal symmetry}
\label{sec:Tbreak}
Consider breaking time reversal symmetry on the TI surface while maintaining the chemical potential at the electronic Dirac node. This induces a mass term $\bar{\Psi}_e\Psi_e$ leading to an insulating surface with surface Hall conductance $\sigma_{xy}=\frac12 \frac{e^2}{h}$. The effective response theory on the surface is given by:
\beq
L = -\frac{1}{2 (4 \pi)}  \epsilon^{\mu \nu \lambda} A_{\mu}\partial_{\nu} A_{\lambda}
\label{Eq:sigmahalf}
\eeq

The dual composite fermions also acquire a mass gap due to breaking of $T$ symmetry via the mass term $\bar{\psi}_{cf} \psi_{cf}$. Integrating out the single Dirac cone of $\psi_{cf}$, leads to the following effective action:

\beq
L = \frac{1}{2 (4\pi)}  \epsilon^{\mu \nu \lambda}  a_{\mu} \partial_{\nu} a_{\lambda} -\frac{1}{2 (2\pi)} \epsilon^{\mu \nu \lambda}  A_{\mu}  \d_{\nu} a_{\lambda}
\eeq
Integrating out the dynamical gauge field $a_{\mu}$ then produces the same Lagrangian as  Eq.~(\ref{Eq:sigmahalf}). 



\section{Parton theories with index $\nu \neq 1$}
\label{sec:othernu}
We have seen in section \ref{sec:bulkpartons} how to obtain a $T$-invariant electronic insulator with $\theta_{EM} = \pi$ by confining the spin-liquid phase SL$_\times$. This spin-liquid phase was obtained through a parton construction, with partons $\psi$ placed into a non-interacting $\nu = 1$ band-structure of class AIII. As we already mentioned, non-interacting electron phases in class AIII have an integer classification $\nu \in {\mathbb Z}$, which collapes to a ${\mathbb Z}_8$ group upon adding interactions. We now ask what happens if we place our partons into a non-interacting band-structure with $\nu \neq 1$ and $\nu$ - odd? 

One can quickly see that the properties of the electronic $\nu = 1$ phase that we used in our bulk construction in section \ref{sec:bulkpartons} are shared by all phases with $\nu$ - odd. In particular, they all have a response to a $u(1)$ gauge field characterized by $\theta = \pi$. The quantum numbers of dyons in the bosonic spin-liquids SL$^{\nu}_\times$ based on a parton band-structure with $\nu$-odd will, therefore, be identical. Consequently, the associated confined phases obtained by condensing the $D$-dyon $\Psi_e \times (0, 2)$ will all have a $\theta_{EM} = \pi$ response to the $U(1)$ gauge-field $A_{\mu}$. The resulting surface theories are, however, different, consisting of $N_f = |\nu|$ flavors of gapless Dirac fermions interacting with a $u(1)$ gauge field $a_{\mu}$. 

We now discuss whether the bulk SL$^{\nu}_{\times}$ phases with different $\nu$ and the associated confined phases are, in fact, different. First of all, due to the collapse of the non-interacting classification, the bulk phase only depends on $\nu$ mod 8. Furthermore, we claim that once the $u(1)$ symmetry of class AIII is gauged,  $\nu = 1$ and  $\nu = -1 \sim 7$ phases are identical. Recall that when the $u(1)$ symmetry is a global symmetry of an electronic theory, $\nu = 1$ and  $\nu = -1$ phases differ in the action of the $T$ symmetry on a single monopole $(q = 1/2, m =1)$. Under $T$, $(1/2, 1) \leftrightarrow (-1/2, 1)$. In a theory of electrons, these two time-reversal partners differ by a local object - the electron $(1, 0)$, and one can, therefore, assign a $T^2$ value to them. In the $\nu = 1$ state, $(1/2, 1)$ has $T^2 = +i$ and $(-1/2, 1)$ has $T^2 = -i$. In the $\nu = -1$ state the $T^2$ assignments of these monopoles are reversed.\cite{MetlitskiChenFidkowskiAV2014} However, once we treat the $u(1)$ symmetry as a gauge-symmetry, the parton $\psi = (1,0)$ is no longer a local object. Therefore, the dyons $(1/2, 1)$ and $(-1/2, 1)$ belong to different topological sectors and one cannot assign a value of $T^2$ to them. Similarly, $\nu = 3$ and $\nu = -3 \sim 5$ phases collapse after gauging. 

It remains to see whether SL$^{\nu = 1}$ and SL$^{\nu = 5}$ bulk phases are distinct. In fact, they are. Recall that before the $u(1)$ symmetry is gauged, $\nu = 1$ and $\nu = 5$ phases differ by the eTmT SPT phase of neutral bosons. But neutral bosons are not affected by the gauging of $u(1)$ symmetry. Therefore, SL$^{\nu = 1}_{\times}$ and SL$^{\nu = 5}_{\times}$ spin-liquid phases also differ by an eTmT phase of neutral bosons. Likewise, the associated confined phases also only differ by an eTmT phase. At the level of confined phases, this can also be seen by considering the symmetric topological surface states in the $\nu = 1$ and $\nu = 5$ constructions. As we discussed in section \ref{sec:surfacephases}, these states can be obtained by pair-condensing the composite fermions $\psi_{cf}$ in the surface QED$_3$ theory. In the case of both $\nu = 1$ and $\nu = 5$ the resulting intrinsic topological order is given by the T-Pfaffian, however, $\nu = 1$ and $\nu = 5$ differ in the action of time-reversal symmetry on the anyons of T-Pfaffian. In the $\nu = 1$ case the charge $e/4$ anyon $\sigma_1$ is a Kramers singlet and the associated state is known as T-Pfaffian$_+$. In the $\nu = 5$ case the charge $e/4$ anyon is a Kramers doublet and the associated state is known as T-Pfaffian$_-$ (see table \ref{TPfaffian}). These two T-Pfaffian states are known to differ precisely by the eTmT surface topological order (i.e. T-Pfaffian$_+$ + eTmT can be driven via a surface phase transition to T-Pfaffian$_-$).

As we already mentioned, since the confined phases have $\theta_{EM} = \pi$ they differ from a non-interacting TI at most by an SPT phase of neutral bosons  with $T$-invariance.\cite{ChongScience} Such $T$-invariant boson SPT phases have a ${\mathbb Z}^2_2$ classification.\cite{AVTS,Burnell2013,KapustinBos, Kapustin2014a} The two $\mathbb{Z}_2$ root phases are best understood via their symmetric topologically ordered surface states. One of the root phases admits {  the aforementioned eTmT surface topological order}. The other root phase admits a surface topological order with anyons $\{1, f_1, f_2, f_3\}$, where $f_1$, $f_2$, $f_3$ are fermions and the fusion rules are the same as in a toric code. This phase (and the above topological order) is abbreviated as fff. Thus, {  our confined phases} are identical to a non-interacting electron TI up to these bosonic SPT phases. In fact, one can rule-out the scenario where the confined phases differ from the ordinary TI by the fff state (or fff + eTmT). Indeed, if one strongly breaks the $T$-symmetry on the surface of an fff state, one drives the surface  into a topologically trivial phase with thermal Hall response $\kappa_{xy}/T  = 4$ and electric Hall response $\sigma_{xy}/T = 0$. However, the trivial $T$-broken surface phase of an ordinary TI has $\sigma_{xy} = \kappa_{xy}/T = 1/2$. Similarly, if we break $T$ strongly starting from the T-Pfaffian$_\pm$ surface states of $\nu = 1$, $\nu = 5$ confined phases, we obtain $\sigma_{xy} = \kappa_{xy}/T = 1/2$. Strictly 2d phases of fermions with no intrinsic topological order always have $\sigma_{xy} - \kappa_{xy}/T \equiv 0$ (mod 8). Therefore, our $\nu = 1$ and $\nu = 5$ confined phases differ from the non-interacting TI at most by the eTmT phase. Since $\nu = 1$ and $\nu = 5$ themselves differ by the eTmT phase, we conclude that one of them is continuosly connected to the TI. By strengthening the arguments presented in section \ref{sec:bulkduality}, one can show that it is actually the $\nu = 1$ phase (surface topological order T-Pfaffian$_+$), which corresponds to the ordinary TI.

From the above discussion, we obtain a family of novel surface theories for the ordinary TI. Since all $\nu$'s of the form $\nu = 8 k \pm 1$ give rise to the same bulk phase, QED$_3$ with $N_f = 8 k \pm 1$ flavors provides a description of the TI surface. Let us start with the weakly coupled QED$_3$ in the UV and ask about its fate in the IR. If one considers the $SU(N_f)$ invariant situation, in the limit $N_f \to \infty$ the IR theory is a CFT. This CFT is under complete theoretical control and one can systematically compute scaling dimensions of operators in powers of $1/N_f$. For instance, the $T$-odd ``mass" operator $\bar{\psi}_{cf} \psi_{cf}$ has scaling dimension  $\Delta_{\bar{\psi} \psi} \approx 2 + \frac{128}{3\pi^2N_f}$.\cite{HermeleFisher} The flux $4\pi$ instanton operator corresponding to the physical electron $\Psi_e$ has scaling dimension $\Delta_{\Psi_e} \approx 0.673 N_f$.\cite{KapustinMonopoles} 
Clearly, the CFTs with large $N_f$ are distinct from a free Dirac cone. 
The strong version of particle-vortex duality discussed in section \ref{sec:dynamics} would require that when $N_f = 1$ the IR CFT  becomes a free Dirac cone. 

\section{Bulk duality}
\label{sec:bulkduality}

In section \ref{sec:bulkpartons} we've constructed a 3+1D SPT phase of electrons with symmetry $U(1)\rtimes T$ and electromagnetic response with $\theta_{EM} = \pi$. As we already mentioned, by general arguments of Ref.~\onlinecite{ChongScience} this phase can differ from the non-interacting TI at most by a bosonic SPT phase with $T$-symmetry. We now give a different argument for this. In the process, we will demonstrate that the particle-vortex duality of 2+1 dimensional Dirac fermions can be understood as a descendent of electromagnetic duality of the 3+1 dimensional $u(1)$ gauge theory.

Our construction in section \ref{sec:bulkpartons} started with a $T$-symmetric spin-liquid phase of neutral bosons SL$^{\nu = 1}_\times$. This phase was obtained by using the parton decomposition (\ref{eq:partonB}), assigning partons $T$-transformations (\ref{eq:Tparton}) resulting in an overall symmetry group $u(1) \times T$, and then placing the partons into a $\nu=1$ band-structure of class AIII.  Now consider a (seemingly) different $T$-symmetric spin-liquid phase of neutral bosons obtained through the decomposition,
\beq B = \tilde{\psi}^{\dagger} \tilde{\Gamma} \tilde{\psi} \eeq
with $\tilde{\psi}$ - a fermionic parton. The decomposition again has a $\tilde{u}(1)$ gauge symmetry,
\beq \tilde{u}(1): \,\, \tilde{\psi}(x) \to e^{i \alpha(x)} \tilde{\psi}(x)\eeq
which will give rise to an emergent gauge field $\tilde{a}_{\mu}$. (We use the tilde superscript to distinguish the present construction from the one in section \ref{sec:bulkpartons}). We assign the parton $\tilde{\psi}$ the following transformation properties under $T$,
\beq T: \quad \tilde{\psi} \to \tilde{U}_T \tilde{\psi} \eeq
with $\tilde{U}_T \tilde{U}^*_T = -1$, so that $T^2 \tilde{\psi} (T^{\dagger})^2 = -\tilde{\psi}$. Since now $T$ does not change the $\tilde{u}(1)$ charge of $\tilde{\psi}$, $\tilde{\psi}$ is a true Kramers doublet. The time-reversal symmetry and the gauge symmetry now do not commute: if $\tilde{u}_\alpha$ is a gauge rotation by a phase $\alpha$, $T \tilde{u}_\alpha T^{\dagger} = \tilde{u}_{-\alpha}$, so  the overall symmetry group is $\tilde{u}(1) \rtimes T$. This symmetry group is the same as for familiar topological insulators in class AII. To complete the construction of the spin-liquid phase, we place the partons $\tilde{\psi}$ into a non-interacting TI bandstructure. We label the resulting spin-liquid SL$_\rtimes$.

We will now argue that the two states SL$_\rtimes$ and SL$^{\nu = 1}_\times$, in fact, belong to the same phase.

Let us first discuss the excitations of the SL$_\rtimes$ state. Integrating the partons out, we obtain an effective action for $\tilde{a}_{\mu}$,
\beq S[\tilde{a}_{\mu}] =  \int d^3 x d t \left( \frac{1}{4 \tilde{e}^2} \tilde{f}_{\mu \nu} \tilde{f}^{\mu \nu} + \frac{\theta}{32 \pi^2}  \epsilon^{\mu \nu \lambda \sigma} \tilde{f}_{\mu \nu} \tilde{f}_{\lambda \sigma}\right)\label{eq:Stildea}\eeq
with $\theta=\pi$ and $\tilde{f}_{\mu \nu} = \d_{\mu} \tilde{a}_{\nu} - \d_{\nu} \tilde{a}_{\mu}$. There is again a topological term in the effective action with $\theta = \pi$. Thus, the spectrum of dyon excitations can again be labeled by electric and magnetic charges $(\tilde{q}, \tilde{m})$ with $\tilde{m}$ - integers and $\tilde{q} - \tilde{m}/2$ - integers. As in the SL$_\times$ phase, the self-statistics of dyons is $(-1)^{(\tilde{q} - \tilde{m}/2)(\tilde{m}+1)}$. Two dyons $(\tilde{q}, \tilde{m})$ and $(\tilde{q}', \tilde{m}')$ experience the usual statistical interaction, with a statistical phase $\exp\left(i (\tilde{q} \tilde{m}' - \tilde{q}' \tilde{m}) \Omega/2\right)$, as well as a $1/r$ Coulomb interaction.

Under time-reversal, $T: (\tilde{q}, \tilde{m} ) \to (\tilde{q}, - \tilde{m})$. Only excitations whose topological sector is not modified by $T$ can be assigned a Kramers parity. In the present case, these are the pure-charge excitations $(q, 0)$. The single parton $\tilde{\psi} = (1, 0)$ is (by construction) a Kramers doublet. 




In order to compare the two spin-liquid phases SL$^{\nu= 1}_\times$ and SL$_\rtimes$ it is convenient to choose the following basis for the lattice of dyon excitations. Starting with the SL$_\times$ case, let us choose as a basis the two dyons: $d_+ = (1/2, 1)$ and $d_- = (-1/2, 1)$. These dyons are both bosons, and have a non-trivial mutual statistical interaction: $d_+$ sees $d_-$ as a charge $(1,0)$ would see a monopole $(0,1)$ at $\theta = 0$. Under $T:\, d_+ \leftrightarrow d_-$. These time reversal partners fuse to a double monopole $(0,2)$, which is a Kramers doublet fermion as is required by the presence of a non-trivial statistical interaction between them. Decomposing a general dyon as ${\cal D} = d^{n_+}_+ d^{n_-}_-$, two dyons with quantum numbers $(n_+, n_-)$ and $(n'_+, n'_-)$ have a static interaction:
\begin{eqnarray} 
E &=& \frac{1}{4\pi r} \left( e^2 q q' + \frac{(2 \pi)^2}{e^2} m m'\right)\nonumber\\
 &=& \frac{1}{4 \pi r} \left( \frac{e^2}{4} (n_+ - n_-) (n'_+ - n'_-)  \right . \nonumber \\ &&\left . + \frac{(2 \pi)^2}{e^2} (n_+ + n_-) (n'_+ + n'_-)\right)
  \label{eq:Coulomb}
 \end{eqnarray}

In the SL$_\rtimes$ case, let us choose a different dyon basis: $\tilde{d}_+ = (1/2, -1)$ and $\tilde{d}_- = (1/2, 1)$. Again, these dyons are both bosons and have a non-trivial mutual statistical interaction: $\tilde{d}_+$ sees $\tilde{d}_-$ as a charge sees a monopole at $\theta = 0$. Furthermore, under $T: \tilde{d}_+ \leftrightarrow \tilde{d}_-$ and these two time-reversal partners fuse to a single charge $(1, 0)$, which is a Kramers doublet fermion. Decomposing a general dyon $\tilde{{\cal D}} = \tilde{d}_+^{\tilde{n}_+} \tilde{d}_-^{\tilde{n}_-}$, two dyons with quantum numbers $(\tilde{n}_+, \tilde{n}_-)$ and $(\tilde{n}'_+, \tilde{n}'_-)$ have a static interaction 

\begin{eqnarray}  
E &=& \frac{1}{4 \pi r} \left( \tilde{e}^2 \tilde{q} \tilde{q}' + \frac{(2 \pi)^2}{\tilde{e}^2} \tilde{m} \tilde{m}'\right) \nonumber\\
 & =& \frac{1}{4 \pi r} \left ( \frac{\tilde{e}^2}{4} (\tilde{n}_+ + \tilde{n}_-) (\tilde{n}'_+ + \tilde{n}'_-) \right . \nonumber \\ 
& & +\left .  \frac{(2 \pi)^2}{\tilde{e}^2} (\tilde{n}_+ - \tilde{n}_-) (\tilde{n}'_+ - \tilde{n}'_-)\right )
\end{eqnarray}

We see that the properties of excitations in the $u(1) \times T$ case and in the $\tilde{u}(1) \rtimes T$ case are the same if we identify $d_{+} \sim \tilde{d}_{+}$, $d_- \sim \tilde{d}_-$ and $e = \frac{4 \pi}{\tilde{e}}$. In fact, this duality is just an element of the general $SL(2, \mathbb{Z})$ duality of the $u(1)$ gauge theory. The only non-trivial fact is that this element of the duality preserves the time-reversal symmetry.

Based on the above discussion, one is tempted to conclude that the two spin-liquids SL$^{\nu = 1}_\times$ and SL$_\rtimes$, in fact, belong to the same 
phase. One caveat is that, in principle, these two phases could differ by a bosonic SPT phase with $T$ symmetry (i.e. eTmT phase or fff phase). Indeed, an ``addition" of such an SPT phase will not alter the properties of the excitations charged under the gauge symmetry.  In fact, as discussed in section \ref{sec:othernu}, the SL$^{\nu = 5}_{\times}$ phase based on a $\nu =5$ band-structure of partons $\psi$ differs from the SL$^{\nu = 1}_{\times}$ phase precisely by an eTmT phase. A priori, it is not clear if SL$_{\rtimes}$ is dual to SL$^{\nu = 1}_{\times}$ or SL$^{\nu = 5}_\times$. In a forthcoming work,\cite{Metlitski_ta} we will argue that the duality is, in fact, between SL$_{\rtimes}$ and SL$^{\nu = 1}_\times$. We will briefly summarize the strategy for showing this in section \ref{sec:RP4}.


Having established the duality between two spin-liquid phases, we proceed to confine these phases and obtain a duality between SPT phases of electrons. To do so, imagine adding a trivial band insulator of electrons to each of the spin-liquid phases. In the SL$_\rtimes$ construction, condense the bound state of the physical electron $\Psi_e$ and the single charge $(1,0) = \tilde{d}_+ \tilde{d}_-$ - i.e. the single fermionic parton $\tilde{\psi}$. This bound state is a Kramers singlet boson, so its condensation does not break the $T$ symmetry. The effect of the condensation is to Higgs the dynamical gauge field, effectively ungauging the TI. Indeed, once $\Psi_e \tilde{\psi}$ is condensed, the parton $\tilde{\psi}$ and the electron $\Psi^{\dagger}_e$ become identified, so the resulting phase is continuously connected to a non-interacting TI.  Now, in the dual SL$^{\nu =1}_\times$ description, the single charge $(1, 0) = \tilde{\psi} = \tilde{d}_+ \tilde{d}_-$ corresponds to the double monopole $(0,2) = d_+ d_-$. Hence, condensing $\Psi_e \tilde{\psi}$ in the SL$_\rtimes$ construction is equivalent to condensing the dyon $D = \Psi_e \times (0, 2)$ in the SL$^{\nu=1}_\times$ construction, which is precisely the confinement transition discussed in section \ref{sec:bulkpartons}. We, therefore, conclude that the state obtained by confining the SL$^{\nu =1}_\times$ spin-liquid is continuously connected to a non-interacting TI.


\subsection{Fixing the eTmT factor in the duality}
\label{sec:RP4}
We would like to show that SL$_\rtimes$ and SL$^{\nu =1}_\times$ spin-liquids are identical as $T$-symmetric bosonic phases; in particular, that they do not differ by either the eTmT or the fff phase.

Typically, to detect an SPT phase with a unitary symmetry $G$ using bulk probes only we must ``weakly gauge" $G$, effectively studying the response of the bulk SPT to fluxes of $G$. In the case when the symmetry $G$ is the time-reversal symmetry it has been suggested that the equivalent of ``weakly gauging" the symmetry is placing the system on a non-orientable manifold.\cite{KapustinBos,Kapustin2014a, KapustinFerm} For instance, the partition function of the eTmT phase on the non-orientable manifold $\mathrm{RP}^4$ is equal to $-1$.\cite{KapustinBos,Kapustin2014a} Thus, we can detect whether two phases differ by the eTmT phase by comparing their partition functions on $\mathrm{RP}^4$. Similarly, the partition function of the fff phase on an arbitrary oriented manifold is given by $(-1)^{\sigma(M)}$, where $\sigma(M)$ is the signature of the manifold $M$.\cite{KapustinBos,Walker12} Thus, we can detect whether two phases differ by the fff phase by comparing their partition functions on  CP$^2$, which has signature $\sigma(\mathrm{CP}^2) = 1$. In the forthcoming work,\cite{Metlitski_ta} we will show that the partition functions of SL$_\rtimes$ and SL$^{\nu=1}_\times$ spin-liquids are equal on both $\mathrm{RP}^4$ and $\mathrm{CP}^2$, provided that the coupling constants of the two gauge theories are related by $e = \frac{4 \pi}{\tilde{e}}$. This supports the proclaimed duality.

\section{Conclusions and Future Directions}
In summary, we have derived a new  description of the surface of an electronic topological insulator, given by QED$_3$ with a single gapless two-component Dirac fermion. We argued that these fermions are related to $2hc/e$ vortices of the electron fluid. QED$_3$  represents a dual description of the surface Dirac electrons at the level of Hilbert spaces, operators and symmetries. The dual description allows us to derive well known surface phases of the TI, and also to derive a previously proposed surface topological order - the T-Pfaffian. 

An interesting question for future research is whether  a ``strong" version of this particle-vortex duality holds,  i.e. whether a dynamical equivalence exists between  $N_f = 1$ QED$_3$ and $N_f = 1$ free Dirac fermion, with $N_f$ denoting the number of fermion flavors of the two component fermion fields. Although the conventional folklore holds that QED$_3$ is unstable at small values of $N_f$, we note that our $N_f = 1$ surface theory has no Chern-Simons term. Therefore, it does not fit into the conventional folklore and perhaps can remain gapless in the IR. If so, information about this conformal field theory may be available from the conformal bootstrap\cite{Rychkov}. Much work on particle vortex dualities (or mirror symmetry) have been on supersymmetric (SUSY) theories, which may appear to be irrelevant to the present discussion. However, in \cite{Grover,PonteLee} it was argued that the critical point between the Dirac surface state of a TI and a surface superconductor  is described by a Wess-Zumino model with emergent ${\mathcal N}=2$ SUSY. Thus far a dual theory of this precise model has not appeared, although closely related models have successfully been dualized \cite{Aharony,KapustinBorokhov}. 

\section{Acknowledgements}
We thank Dam Son for inspiring discussions. AV thanks Anton Kapustin, Shamit Kachru and Nathan Seiberg for discussions on S-duality. We would like to thank the organizers of the  Simons Symposium on Quantum Entanglement, where this work was initiated. AV was supported by NSF-DMR 1206728. This research was supported in part by the National Science Foundation under Grant No. NSF PHY11-25915. After completion of this work, Ref. \onlinecite{Wang2015} by  Wang and Senthil appeared, which also points out {  the duality between $u(1)$ gauge theories discussed in section \ref{sec:bulkduality}.} The same authors have also communicated to us a forthcoming preprint with independently derived results on the dual surface theory which agree with our work. We thank them for sharing their results with us. 

\appendix

\section{Band Structure of Class AIII Topological Superconductor}
\label{sec:AIII}

Here we write down an explicit band structure for fermions in class AIII topological superconductor phase. Consider a cubic lattice model with four orbitals per site labelled by $\tau_z=\pm1$ and $\nu_z=\pm 1$. Consider the 1-particle Bloch Hamiltonian:

\begin{eqnarray*}
H_0 &=& t\left [ \sin k_x \alpha_x + \sin k_y \alpha_y+\sin k_z \alpha_z \right ] \\&&+ m \left [  \lambda -(\cos k_x+\cos k_y +\cos k_z) \right ]\beta_5 
\end{eqnarray*}
We have $x_ix_j+x_jx_i = 2\delta_{ij}$ where $x_i\in (\alpha_x,\,\alpha_y,\,\alpha_z,\,\beta_0,\,\beta_5)$. An explicit representation is $(\alpha_x,\, \alpha_y,\,\alpha_z,\,\beta_0,\,\beta_5) = (\tau_x,\,\tau_z\nu_x,\,\tau_z\nu_z,\,\tau_y,\,\tau_z\nu_y)$. This Hamiltonian has a chiral symmetry $\beta_0 H_0\beta_0 = -H_0$.  Time reversal symmetry in the second quantized representation takes the form: $\psi \rightarrow \beta_0 \psi^\dagger$ (and, being an antiunitary symmetry $i\rightarrow -i$). In contrast to regular time reversal symmetry particles are taken to holes, so the conserved U(1) is like spin rather than charge.   

For $\lambda >3$ the model is in a trivial phase. However, for $1<\lambda<3$ the sign of the mass term changes sign at the origin in momentum space indicating this is a $\nu=1$ topological phase in the AIII class. In order to access SL$_\times$, the gauged topological superconductor, we require the fermionic partons to take up a band structure with the same topology.

\section{Compactness of the gauge field in the surface theory}
\label{App:Compactness}
It is often stated that a single Dirac fermion in $2+1$ dimensions suffers from the parity anomaly, namely it cannot be consistently coupled to a $U(1)$ gauge field preserving the time-reversal symmetry.\cite{Redlich} When the Dirac fermion appears as the surface state of a $3+1$ dimensional insulator (either the ordinary TI in class AII with symmetry $U(1) \rtimes T$ or class AIII with symmetry $U(1) \times T$) this anomaly has a well-known resolution: when one gauges the $U(1)$ symmetry, the $U(1)$ gauge-field lives in the $3+1$ dimensional bulk, and the $\theta = \pi$ bulk EM response ``cancels" the anomaly of the surface. Now, our surface theory (\ref{eq:DiracgaugedA}) has a dynamical $u(1)$ gauge field $a_{\mu}$, which is confined to live just on the surface (we switch off the background electromagnetic field $A_{\mu}$ for now). Thus, the standard argument for the evasion of the anomaly via the $3+1$ dimensional bulk does not directly apply in this case. Rather, the anomaly is resolved by modifying the compactification of the gauge-field $a_{\mu}$ in the surface theory. For instance, we already saw that only configurations of $a_{\mu}$ with magnetic flux $2 \pi m$ with $m$-even are allowed on the surface. There is a related restriction on the electric fluxes that one can place through the space-time 2-cycles of the surface.

As an example, imagine that our $D$-condensed phase occupies a solid torus, so that its boundary is a torus $T^2$, with periodic $x$ and $y$ direction of length $L$. We will choose the $x$ cycle to wrap the hole of the solid torus, while the $y$ cycle can be contracted within the solid torus. To simplify the discussion, let us break the $T$-symmetry on the surface by adding a mass term $m \bar{\psi}_{cf} \psi_{cf}$ to the surface theory. The low-energy surface action then becomes,
\beq L = \frac{i k}{4\pi} \epsilon_{\mu \nu \lambda} a_{\mu} \d_{\nu} a_\lambda \label{eq:CS}\eeq
with $k = \frac{1}{2} \mathrm{sgn}(m)$ (for definiteness, let us choose $m > 0$ so that $k = 1/2$). It is a standard statement that the level $k$ of the $2+1$ dimensional Chern-Simons theory must be an integer, which seems inconsistent with our finding of $k = 1/2$. Let us recall what this statement is based on. Let's integrate out the temporal component of the gauge field $a_\tau$. This enforces the constraint $\d_{x} a_y - \d_y a_x = 0$. Then $a_{i}(\vec{x}, \tau) = \d_i \alpha(\vec{x}, \tau) + \frac{\theta_i(\tau)}{L}$, so that the only remaining physical degrees of freedom are $\theta_1$ and $\theta_2$ corresponding to the flux of $a$ through the $x$ and $y$ 1-cycles. The effective action then takes the form,
\beq L = -\frac{i k}{2 \pi} \theta_1 \d_\tau \theta_2 \label{eq:CStheta} \eeq
Now, in a standard $2+1$ dimensional theory, large gauge-transformations, $a_1 \to a_1 + \frac{2 \pi}{L}$, $a_2 \to a_2 + \frac{2 \pi}{L}$ are allowed, corresponding to $\theta_1 \to \theta_1 + 2\pi$, $\theta_2 \to \theta_2 + 2 \pi$. In the path-integral treatment these transformations are implemented by allowing $\theta_{1,2}$ to wind by $2 \pi$ around the temporal circle: $\theta_{i}(\beta) = \theta_{i}(0) + 2\pi n_i$, with $n_i$ - integers. This corresponds to placing electric fluxes $2 \pi n_i$ through the space-time 2-cycles of the system. Now, imagine there is an electric flux $2 \pi$ through the $\tau-y$ cycle. We see that in this case the action (\ref{eq:CS}) changes by $S \to S - 2 \pi i k$ as we shift $\theta_1 \to \theta_1 + 2 \pi$. Thus, the partition function remains invariant only if $k$ is an integer. In particular, for $k = 1/2$ the partition function acquires a phase $-1$. While we have demonstrated this effect in the $T$-broken surface theory (\ref{eq:CS}), it is also present for the $T$-invariant gapless Dirac fermion.\cite{Redlich} 

One encounters the same difficulty if one attempts to quantize the theory (\ref{eq:CStheta}) in real-time. The commutation relation between $\theta_1$ and $\theta_2$ reads,
\beq \left[\theta_1, \theta_2\right] = -\frac{2 \pi i}{k}\eeq
The operators $U_{1,2}$ which implement the large-gauge transformations $\theta_{1,2} \to \theta_{1,2} + 2 \pi$ read $U_1 = e^{-i k \theta_2}$, $U_2 = e^{i k \theta_1}$. Now,
\beq U_1 U_2 = e^{2 \pi i k} U_2 U_1 \eeq
Thus, the large gauge transformations along the two directions commute only if $k$ is an integer. In our theory with $k =1/2$, $U_1$ and $U_2$ anti-commute.

The above anomaly is resolved in our surface theory in the following way. While large gauge transformation $U_1$ shifting $\theta_1 \to \theta_1 + 2\pi$ is allowed, only the transformation $U^2_2$, shifting $\theta_2 \to \theta_2 + 4 \pi$ is permitted. Thus, we are only allowed to place electric flux $2 \pi m$ with $m$ - even along the $y-\tau$ 2-cycle, while a flux $2 \pi m$ with arbitrary integer $m$ can be placed along the $x-\tau$ 2-cycle. Note that from the bulk point of view the two cycles are not equivalent: the $x$ cycle is uncontractible in the 3d solid torus, while the $y$ cycle is contractible. Now, $U_1$ and $U^2_2$ commute and we can compute the ground state degeneracy. Working in the $\theta_1$ basis and imposing $U^2_2 = 1$ we must have $\theta_1 = 2 \pi l$ with $l$ - integer. Since $\theta_1$ is identified modulo $2 \pi$, we have a unique physical ground state given by $\theta_1 = 0$. This is consistent with our expectations. Indeed, the $T$-broken surface state has no intrinsic topological order so it should possess no ground state degeneracy on a torus. The only excitation is the gapped $\psi_{cf}$. The Chern-Simons field (\ref{eq:CS}) attaches flux $4\pi$ to $\psi_{cf}$, which preserves its fermionic statistics. Now, a flux $4 \pi$ instanton will create $\psi_{cf}$ with an attached flux $4 \pi$. Recalling that a flux $4\pi$ instanton corresponds to the electron creation operator, in the $T$-broken phase $\psi_{cf}$ is identified with the electron.

We can directly understand the restriction on the allowed large gauge transformations of the surface theory using our bulk construction. Let us imagine a process where $\theta_2(\tau) = 2 \pi \tau/\beta$, i.e. $\theta_2$ winds by $2 \pi$ along the temporal cycle. This gives rise to an electric field $\vec{e}  = (0, \frac{2 \pi}{\beta L})$ along the surface. As discussed in the previous section this electric field will be Meissner screened by a current of $D$-dyons along the surface with surface density, $j^D_i = -\frac{1}{4 \pi} \epsilon_{ij} e_j = (-\frac{1}{2} \frac{1}{\beta L},0)$. Note that the dyon current is along the uncontractible $x$-cycle of the solid torus. The total number of $D$-dyons that  have passed through the $x = 0$ cross-section of the solid torus in the time $0 < \tau < \beta$ is $N_D = -1/2$. Now, if the system at time $\tau = \beta$ comes back to its initial $\tau = 0$ configuration then an integer number of $D$-dyons must have passed through the $x = 0$ cross-section. Therefore, we conclude that the system has not returned to its initial configuration at $\tau = \beta$. Therefore, $\theta_2 = 0$ and $\theta_2 = 2 \pi$ are not indentical, rather $\theta_2 \sim \theta_2 + 4\pi$. On the other hand, $\theta_1$ is, indeed, periodic modulo $2\pi$. Indeed, when an electric field is applied along the $x$ direction, the dyons move along the $y$ cycle. Since this cycle is contractible, the number of dyons that pass through any cross-section of the solid torus is now zero. Thus, $\theta_1 \sim \theta_1 + 2\pi$.

We conclude that the surface QED$_3$ theory differs from the conventional $2+1$-dimensional $u(1)$ gauge theory in the allowed large gauge transformations. Once the set of large gauge transformations is restricted, QED$_3$ with a single Dirac cone becomes fully consistent with $T$-symmetry.

\bibliography{Refs}

\begin{thebibliography}{51}%
\makeatletter
\providecommand \@ifxundefined [1]{%
 \@ifx{#1\undefined}
}%
\providecommand \@ifnum [1]{%
 \ifnum #1\expandafter \@firstoftwo
 \else \expandafter \@secondoftwo
 \fi
}%
\providecommand \@ifx [1]{%
 \ifx #1\expandafter \@firstoftwo
 \else \expandafter \@secondoftwo
 \fi
}%
\providecommand \natexlab [1]{#1}%
\providecommand \enquote  [1]{``#1''}%
\providecommand \bibnamefont  [1]{#1}%
\providecommand \bibfnamefont [1]{#1}%
\providecommand \citenamefont [1]{#1}%
\providecommand \href@noop [0]{\@secondoftwo}%
\providecommand \href [0]{\begingroup \@sanitize@url \@href}%
\providecommand \@href[1]{\@@startlink{#1}\@@href}%
\providecommand \@@href[1]{\endgroup#1\@@endlink}%
\providecommand \@sanitize@url [0]{\catcode `\\12\catcode `\$12\catcode
  `\&12\catcode `\#12\catcode `\^12\catcode `\_12\catcode `\%12\relax}%
\providecommand \@@startlink[1]{}%
\providecommand \@@endlink[0]{}%
\providecommand \url  [0]{\begingroup\@sanitize@url \@url }%
\providecommand \@url [1]{\endgroup\@href {#1}{\urlprefix }}%
\providecommand \urlprefix  [0]{URL }%
\providecommand \Eprint [0]{\href }%
\providecommand \doibase [0]{http://dx.doi.org/}%
\providecommand \selectlanguage [0]{\@gobble}%
\providecommand \bibinfo  [0]{\@secondoftwo}%
\providecommand \bibfield  [0]{\@secondoftwo}%
\providecommand \translation [1]{[#1]}%
\providecommand \BibitemOpen [0]{}%
\providecommand \bibitemStop [0]{}%
\providecommand \bibitemNoStop [0]{.\EOS\space}%
\providecommand \EOS [0]{\spacefactor3000\relax}%
\providecommand \BibitemShut  [1]{\csname bibitem#1\endcsname}%
\let\auto@bib@innerbib\@empty
\bibitem [{\citenamefont {Franz}\ and\ \citenamefont
  {Molenkamp}(2013)}]{FranzMolenkamp2013}%
  \BibitemOpen
  \bibinfo {editor} {\bibfnamefont {M.}~\bibnamefont {Franz}}\ and\ \bibinfo
  {editor} {\bibfnamefont {L.}~\bibnamefont {Molenkamp}},\ eds.,\ \href@noop {}
  {\emph {\bibinfo {title} {Topological Insulators}}},\ \bibinfo {series}
  {Contemporary Concepts of Condensed Matter Science}, Vol.~\bibinfo {volume}
  {6}\ (\bibinfo  {publisher} {Elsevier},\ \bibinfo {year} {2013})\BibitemShut
  {NoStop}%
\bibitem [{\citenamefont {Haldane}(1983)}]{Haldane}%
  \BibitemOpen
  \bibfield  {author} {\bibinfo {author} {\bibfnamefont {F.~D.~M.}\
  \bibnamefont {Haldane}},\ }\href@noop {} {\bibfield  {journal} {\bibinfo
  {journal} {Phys. Rev. Lett.}\ }\textbf {\bibinfo {volume} {50}},\ \bibinfo
  {pages} {1153} (\bibinfo {year} {1983})}\BibitemShut {NoStop}%
\bibitem [{\citenamefont {{Chen}}\ \emph {et~al.}(2011)\citenamefont {{Chen}},
  \citenamefont {{Gu}},\ and\ \citenamefont {{Wen}}}]{Chen1d}%
  \BibitemOpen
  \bibfield  {author} {\bibinfo {author} {\bibfnamefont {X.}~\bibnamefont
  {{Chen}}}, \bibinfo {author} {\bibfnamefont {Z.-C.}\ \bibnamefont {{Gu}}}, \
  and\ \bibinfo {author} {\bibfnamefont {X.-G.}\ \bibnamefont {{Wen}}},\
  }\href@noop {} {\bibfield  {journal} {\bibinfo  {journal} {Phys. Rev. B}\
  }\textbf {\bibinfo {volume} {83}},\ \bibinfo {eid} {035107} (\bibinfo {year}
  {2011})}\BibitemShut {NoStop}%
\bibitem [{\citenamefont {Fidkowski}\ and\ \citenamefont
  {Kitaev}(2010)}]{Fidkowski0}%
  \BibitemOpen
  \bibfield  {author} {\bibinfo {author} {\bibfnamefont {L.}~\bibnamefont
  {Fidkowski}}\ and\ \bibinfo {author} {\bibfnamefont {A.}~\bibnamefont
  {Kitaev}},\ }\href@noop {} {\bibfield  {journal} {\bibinfo  {journal} {Phys.
  Rev. B}\ }\textbf {\bibinfo {volume} {81}},\ \bibinfo {pages} {134509}
  (\bibinfo {year} {2010})}\BibitemShut {NoStop}%
\bibitem [{\citenamefont {Fidkowski}\ and\ \citenamefont
  {Kitaev}(2011)}]{Fidkowski1}%
  \BibitemOpen
  \bibfield  {author} {\bibinfo {author} {\bibfnamefont {L.}~\bibnamefont
  {Fidkowski}}\ and\ \bibinfo {author} {\bibfnamefont {A.}~\bibnamefont
  {Kitaev}},\ }\href@noop {} {\bibfield  {journal} {\bibinfo  {journal} {Phys.
  Rev. B}\ }\textbf {\bibinfo {volume} {83}},\ \bibinfo {pages} {075103}
  (\bibinfo {year} {2011})}\BibitemShut {NoStop}%
\bibitem [{\citenamefont {Turner}\ \emph {et~al.}(2011)\citenamefont {Turner},
  \citenamefont {Pollmann},\ and\ \citenamefont {Berg}}]{Turner1d}%
  \BibitemOpen
  \bibfield  {author} {\bibinfo {author} {\bibfnamefont {A.~M.}\ \bibnamefont
  {Turner}}, \bibinfo {author} {\bibfnamefont {F.}~\bibnamefont {Pollmann}}, \
  and\ \bibinfo {author} {\bibfnamefont {E.}~\bibnamefont {Berg}},\ }\href@noop
  {} {\bibfield  {journal} {\bibinfo  {journal} {Phys. Rev. B}\ }\textbf
  {\bibinfo {volume} {83}},\ \bibinfo {pages} {075102} (\bibinfo {year}
  {2011})}\BibitemShut {NoStop}%
\bibitem [{\citenamefont {Levin}\ and\ \citenamefont {Gu}(2012)}]{LevinGu}%
  \BibitemOpen
  \bibfield  {author} {\bibinfo {author} {\bibfnamefont {M.}~\bibnamefont
  {Levin}}\ and\ \bibinfo {author} {\bibfnamefont {Z.-C.}\ \bibnamefont {Gu}},\
  }\href@noop {} {\bibfield  {journal} {\bibinfo  {journal} {Phys. Rev. B}\
  }\textbf {\bibinfo {volume} {86}},\ \bibinfo {pages} {115109} (\bibinfo
  {year} {2012})}\BibitemShut {NoStop}%
\bibitem [{\citenamefont {Lu}\ and\ \citenamefont
  {Vishwanath}(2012)}]{LuAV_SPT}%
  \BibitemOpen
  \bibfield  {author} {\bibinfo {author} {\bibfnamefont {Y.-M.}\ \bibnamefont
  {Lu}}\ and\ \bibinfo {author} {\bibfnamefont {A.}~\bibnamefont
  {Vishwanath}},\ }\href@noop {} {\bibfield  {journal} {\bibinfo  {journal}
  {Phys. Rev. B}\ }\textbf {\bibinfo {volume} {86}},\ \bibinfo {pages} {125119}
  (\bibinfo {year} {2012})}\BibitemShut {NoStop}%
\bibitem [{\citenamefont {Vishwanath}\ and\ \citenamefont
  {Senthil}(2013)}]{AVTS}%
  \BibitemOpen
  \bibfield  {author} {\bibinfo {author} {\bibfnamefont {A.}~\bibnamefont
  {Vishwanath}}\ and\ \bibinfo {author} {\bibfnamefont {T.}~\bibnamefont
  {Senthil}},\ }\href@noop {} {\bibfield  {journal} {\bibinfo  {journal} {Phys.
  Rev. X}\ }\textbf {\bibinfo {volume} {3}},\ \bibinfo {pages} {011016}
  (\bibinfo {year} {2013})}\BibitemShut {NoStop}%
\bibitem [{\citenamefont {Metlitski}\ \emph
  {et~al.}(2013{\natexlab{a}})\citenamefont {Metlitski}, \citenamefont {Kane},\
  and\ \citenamefont {Fisher}}]{MetlitskibTI}%
  \BibitemOpen
  \bibfield  {author} {\bibinfo {author} {\bibfnamefont {M.~A.}\ \bibnamefont
  {Metlitski}}, \bibinfo {author} {\bibfnamefont {C.~L.}\ \bibnamefont {Kane}},
  \ and\ \bibinfo {author} {\bibfnamefont {M.~P.~A.}\ \bibnamefont {Fisher}},\
  }\href@noop {} {\bibfield  {journal} {\bibinfo  {journal} {Phys. Rev. B}\
  }\textbf {\bibinfo {volume} {88}},\ \bibinfo {pages} {035131} (\bibinfo
  {year} {2013}{\natexlab{a}})}\BibitemShut {NoStop}%
\bibitem [{\citenamefont {Bi}\ \emph {et~al.}(2013)\citenamefont {Bi},
  \citenamefont {Rasmussen},\ and\ \citenamefont {Xu}}]{Bi2013}%
  \BibitemOpen
  \bibfield  {author} {\bibinfo {author} {\bibfnamefont {Z.}~\bibnamefont
  {Bi}}, \bibinfo {author} {\bibfnamefont {A.}~\bibnamefont {Rasmussen}}, \
  and\ \bibinfo {author} {\bibfnamefont {C.}~\bibnamefont {Xu}},\ }\href@noop
  {} {\bibfield  {journal} {\bibinfo  {journal} {ArXiv e-prints 1309.0515}\ }
  (\bibinfo {year} {2013})},\ \Eprint {http://arxiv.org/abs/1309.0515}
  {arXiv:1309.0515 [cond-mat.str-el]} \BibitemShut {NoStop}%
\bibitem [{\citenamefont {Kitaev}()}]{Kitaev_pc}%
  \BibitemOpen
  \bibfield  {author} {\bibinfo {author} {\bibfnamefont {A.}~\bibnamefont
  {Kitaev}},\ }\href@noop {} {}\bibinfo {howpublished}
  {unpublished}\BibitemShut {NoStop}%
\bibitem [{\citenamefont {Kapustin}\ and\ \citenamefont
  {Thorngren}(2014)}]{Kapustin2014}%
  \BibitemOpen
  \bibfield  {author} {\bibinfo {author} {\bibfnamefont {A.}~\bibnamefont
  {Kapustin}}\ and\ \bibinfo {author} {\bibfnamefont {R.}~\bibnamefont
  {Thorngren}},\ }\href@noop {} {\bibfield  {journal} {\bibinfo  {journal}
  {ArXiv e-prints 1403.0617}\ } (\bibinfo {year} {2014})},\ \Eprint
  {http://arxiv.org/abs/1403.0617} {arXiv:1403.0617 [cond-mat.str-el]}
  \BibitemShut {NoStop}%
\bibitem [{\citenamefont {{Kapustin}}(2014)}]{Kapustin2014a}%
  \BibitemOpen
  \bibfield  {author} {\bibinfo {author} {\bibfnamefont {A.}~\bibnamefont
  {{Kapustin}}},\ }\href@noop {} {\bibfield  {journal} {\bibinfo  {journal}
  {ArXiv e-prints 1404.6659}\ } (\bibinfo {year} {2014})},\ \Eprint
  {http://arxiv.org/abs/1404.6659} {arXiv:1404.6659 [cond-mat.str-el]}
  \BibitemShut {NoStop}%
\bibitem [{\citenamefont {{Wang}}\ \emph {et~al.}(2014)\citenamefont {{Wang}},
  \citenamefont {{Potter}},\ and\ \citenamefont {{Senthil}}}]{ChongScience}%
  \BibitemOpen
  \bibfield  {author} {\bibinfo {author} {\bibfnamefont {C.}~\bibnamefont
  {{Wang}}}, \bibinfo {author} {\bibfnamefont {A.~C.}\ \bibnamefont
  {{Potter}}}, \ and\ \bibinfo {author} {\bibfnamefont {T.}~\bibnamefont
  {{Senthil}}},\ }\href@noop {} {\bibfield  {journal} {\bibinfo  {journal}
  {Science}\ }\textbf {\bibinfo {volume} {343}},\ \bibinfo {pages} {629}
  (\bibinfo {year} {2014})},\ \Eprint {http://arxiv.org/abs/1306.3238}
  {arXiv:1306.3238 [cond-mat.str-el]} \BibitemShut {NoStop}%
\bibitem [{\citenamefont {{Senthil}}(2014)}]{Senthil2014}%
  \BibitemOpen
  \bibfield  {author} {\bibinfo {author} {\bibfnamefont {T.}~\bibnamefont
  {{Senthil}}},\ }\href@noop {} {\bibfield  {journal} {\bibinfo  {journal}
  {ArXiv e-prints 1405.4015}\ } (\bibinfo {year} {2014})},\ \Eprint
  {http://arxiv.org/abs/1405.4015} {arXiv:1405.4015 [cond-mat.str-el]}
  \BibitemShut {NoStop}%
\bibitem [{\citenamefont {Fidkowski}\ \emph {et~al.}(2013)\citenamefont
  {Fidkowski}, \citenamefont {Chen},\ and\ \citenamefont
  {Vishwanath}}]{FidkowskiChenAV}%
  \BibitemOpen
  \bibfield  {author} {\bibinfo {author} {\bibfnamefont {L.}~\bibnamefont
  {Fidkowski}}, \bibinfo {author} {\bibfnamefont {X.}~\bibnamefont {Chen}}, \
  and\ \bibinfo {author} {\bibfnamefont {A.}~\bibnamefont {Vishwanath}},\
  }\href {\doibase 10.1103/PhysRevX.3.041016} {\bibfield  {journal} {\bibinfo
  {journal} {Phys. Rev. X}\ }\textbf {\bibinfo {volume} {3}},\ \bibinfo {pages}
  {041016} (\bibinfo {year} {2013})}\BibitemShut {NoStop}%
\bibitem [{\citenamefont {Wang}\ and\ \citenamefont
  {Senthil}(2014)}]{Wang2014}%
  \BibitemOpen
  \bibfield  {author} {\bibinfo {author} {\bibfnamefont {C.}~\bibnamefont
  {Wang}}\ and\ \bibinfo {author} {\bibfnamefont {T.}~\bibnamefont {Senthil}},\
  }\href@noop {} {\bibfield  {journal} {\bibinfo  {journal} {Phys. Rev. B}\
  }\textbf {\bibinfo {volume} {89}},\ \bibinfo {pages} {195124} (\bibinfo
  {year} {2014})}\BibitemShut {NoStop}%
\bibitem [{\citenamefont {{Metlitski}}\ \emph {et~al.}(2014)\citenamefont
  {{Metlitski}}, \citenamefont {{Fidkowski}}, \citenamefont {{Chen}},\ and\
  \citenamefont {{Vishwanath}}}]{MetlitskiChenFidkowskiAV2014}%
  \BibitemOpen
  \bibfield  {author} {\bibinfo {author} {\bibfnamefont {M.~A.}\ \bibnamefont
  {{Metlitski}}}, \bibinfo {author} {\bibfnamefont {L.}~\bibnamefont
  {{Fidkowski}}}, \bibinfo {author} {\bibfnamefont {X.}~\bibnamefont {{Chen}}},
  \ and\ \bibinfo {author} {\bibfnamefont {A.}~\bibnamefont {{Vishwanath}}},\
  }\href@noop {} {\bibfield  {journal} {\bibinfo  {journal} {ArXiv e-prints}\ }
  (\bibinfo {year} {2014})},\ \Eprint {http://arxiv.org/abs/1406.3032}
  {arXiv:1406.3032 [cond-mat.str-el]} \BibitemShut {NoStop}%
\bibitem [{\citenamefont {{You}}\ \emph {et~al.}(2014)\citenamefont {{You}},
  \citenamefont {{BenTov}},\ and\ \citenamefont {{Xu}}}]{You2014}%
  \BibitemOpen
  \bibfield  {author} {\bibinfo {author} {\bibfnamefont {Y.}~\bibnamefont
  {{You}}}, \bibinfo {author} {\bibfnamefont {Y.}~\bibnamefont {{BenTov}}}, \
  and\ \bibinfo {author} {\bibfnamefont {C.}~\bibnamefont {{Xu}}},\ }\href@noop
  {} {\bibfield  {journal} {\bibinfo  {journal} {ArXiv e-prints}\ } (\bibinfo
  {year} {2014})},\ \Eprint {http://arxiv.org/abs/1402.4151} {arXiv:1402.4151
  [cond-mat.str-el]} \BibitemShut {NoStop}%
\bibitem [{\citenamefont {Dzero}\ \emph {et~al.}(2010)\citenamefont {Dzero},
  \citenamefont {Sun}, \citenamefont {Galitski},\ and\ \citenamefont
  {Coleman}}]{Coleman2010}%
  \BibitemOpen
  \bibfield  {author} {\bibinfo {author} {\bibfnamefont {M.}~\bibnamefont
  {Dzero}}, \bibinfo {author} {\bibfnamefont {K.}~\bibnamefont {Sun}}, \bibinfo
  {author} {\bibfnamefont {V.}~\bibnamefont {Galitski}}, \ and\ \bibinfo
  {author} {\bibfnamefont {P.}~\bibnamefont {Coleman}},\ }\href {\doibase
  10.1103/PhysRevLett.104.106408} {\bibfield  {journal} {\bibinfo  {journal}
  {Phys. Rev. Lett.}\ }\textbf {\bibinfo {volume} {104}},\ \bibinfo {pages}
  {106408} (\bibinfo {year} {2010})}\BibitemShut {NoStop}%
\bibitem [{\citenamefont {Wang}\ and\ \citenamefont
  {Senthil}(2013)}]{Wang2013}%
  \BibitemOpen
  \bibfield  {author} {\bibinfo {author} {\bibfnamefont {C.}~\bibnamefont
  {Wang}}\ and\ \bibinfo {author} {\bibfnamefont {T.}~\bibnamefont {Senthil}},\
  }\href@noop {} {\bibfield  {journal} {\bibinfo  {journal} {Phys. Rev. B}\
  }\textbf {\bibinfo {volume} {87}},\ \bibinfo {pages} {235122} (\bibinfo
  {year} {2013})}\BibitemShut {NoStop}%
\bibitem [{\citenamefont {{Burnell}}\ \emph {et~al.}(2013)\citenamefont
  {{Burnell}}, \citenamefont {{Chen}}, \citenamefont {{Fidkowski}},\ and\
  \citenamefont {{Vishwanath}}}]{Burnell2013}%
  \BibitemOpen
  \bibfield  {author} {\bibinfo {author} {\bibfnamefont {F.~J.}\ \bibnamefont
  {{Burnell}}}, \bibinfo {author} {\bibfnamefont {X.}~\bibnamefont {{Chen}}},
  \bibinfo {author} {\bibfnamefont {L.}~\bibnamefont {{Fidkowski}}}, \ and\
  \bibinfo {author} {\bibfnamefont {A.}~\bibnamefont {{Vishwanath}}},\
  }\href@noop {} {\bibfield  {journal} {\bibinfo  {journal} {ArXiv e-prints}\ }
  (\bibinfo {year} {2013})},\ \Eprint {http://arxiv.org/abs/1302.7072}
  {arXiv:1302.7072 [cond-mat.str-el]} \BibitemShut {NoStop}%
\bibitem [{\citenamefont {Bonderson}\ \emph {et~al.}(2013)\citenamefont
  {Bonderson}, \citenamefont {Nayak},\ and\ \citenamefont
  {Qi}}]{Bonderson2013}%
  \BibitemOpen
  \bibfield  {author} {\bibinfo {author} {\bibfnamefont {P.}~\bibnamefont
  {Bonderson}}, \bibinfo {author} {\bibfnamefont {C.}~\bibnamefont {Nayak}}, \
  and\ \bibinfo {author} {\bibfnamefont {X.-L.}\ \bibnamefont {Qi}},\
  }\href@noop {} {\bibfield  {journal} {\bibinfo  {journal} {Journal of
  Statistical Mechanics: Theory and Experiment}\ }\textbf {\bibinfo {volume}
  {2013}},\ \bibinfo {pages} {P09016} (\bibinfo {year} {2013})}\BibitemShut
  {NoStop}%
\bibitem [{\citenamefont {Metlitski}\ \emph
  {et~al.}(2013{\natexlab{b}})\citenamefont {Metlitski}, \citenamefont {Kane},\
  and\ \citenamefont {Fisher}}]{Metlitski2013}%
  \BibitemOpen
  \bibfield  {author} {\bibinfo {author} {\bibfnamefont {M.~A.}\ \bibnamefont
  {Metlitski}}, \bibinfo {author} {\bibfnamefont {C.~L.}\ \bibnamefont {Kane}},
  \ and\ \bibinfo {author} {\bibfnamefont {M.~P.~A.}\ \bibnamefont {Fisher}},\
  }\href@noop {} {\bibfield  {journal} {\bibinfo  {journal} {ArXiv e-prints
  1306.3286}\ } (\bibinfo {year} {2013}{\natexlab{b}})},\ \Eprint
  {http://arxiv.org/abs/1306.3286} {arXiv:1306.3286 [cond-mat.str-el]}
  \BibitemShut {NoStop}%
\bibitem [{\citenamefont {Wang}\ \emph {et~al.}(2013)\citenamefont {Wang},
  \citenamefont {Potter},\ and\ \citenamefont {Senthil}}]{Wang2013a}%
  \BibitemOpen
  \bibfield  {author} {\bibinfo {author} {\bibfnamefont {C.}~\bibnamefont
  {Wang}}, \bibinfo {author} {\bibfnamefont {A.~C.}\ \bibnamefont {Potter}}, \
  and\ \bibinfo {author} {\bibfnamefont {T.}~\bibnamefont {Senthil}},\ }\href
  {\doibase 10.1103/PhysRevB.88.115137} {\bibfield  {journal} {\bibinfo
  {journal} {Phys. Rev. B}\ }\textbf {\bibinfo {volume} {88}},\ \bibinfo
  {pages} {115137} (\bibinfo {year} {2013})}\BibitemShut {NoStop}%
\bibitem [{\citenamefont {Chen}\ \emph {et~al.}(2014)\citenamefont {Chen},
  \citenamefont {Fidkowski},\ and\ \citenamefont {Vishwanath}}]{Chen2014PRB}%
  \BibitemOpen
  \bibfield  {author} {\bibinfo {author} {\bibfnamefont {X.}~\bibnamefont
  {Chen}}, \bibinfo {author} {\bibfnamefont {L.}~\bibnamefont {Fidkowski}}, \
  and\ \bibinfo {author} {\bibfnamefont {A.}~\bibnamefont {Vishwanath}},\
  }\href {\doibase 10.1103/PhysRevB.89.165132} {\bibfield  {journal} {\bibinfo
  {journal} {Phys. Rev. B}\ }\textbf {\bibinfo {volume} {89}},\ \bibinfo
  {pages} {165132} (\bibinfo {year} {2014})}\BibitemShut {NoStop}%
\bibitem [{\citenamefont {Redlich}(1984)}]{Redlich}%
  \BibitemOpen
  \bibfield  {author} {\bibinfo {author} {\bibfnamefont {A.~N.}\ \bibnamefont
  {Redlich}},\ }\href {\doibase 10.1103/PhysRevLett.52.18} {\bibfield
  {journal} {\bibinfo  {journal} {Phys. Rev. Lett.}\ }\textbf {\bibinfo
  {volume} {52}},\ \bibinfo {pages} {18} (\bibinfo {year} {1984})}\BibitemShut
  {NoStop}%
\bibitem [{\citenamefont {Niemi}\ and\ \citenamefont
  {Semenoff}(1983)}]{Semenoff}%
  \BibitemOpen
  \bibfield  {author} {\bibinfo {author} {\bibfnamefont {A.~J.}\ \bibnamefont
  {Niemi}}\ and\ \bibinfo {author} {\bibfnamefont {G.~W.}\ \bibnamefont
  {Semenoff}},\ }\href {\doibase 10.1103/PhysRevLett.51.2077} {\bibfield
  {journal} {\bibinfo  {journal} {Phys. Rev. Lett.}\ }\textbf {\bibinfo
  {volume} {51}},\ \bibinfo {pages} {2077} (\bibinfo {year}
  {1983})}\BibitemShut {NoStop}%
\bibitem [{\citenamefont {Halperin}\ \emph {et~al.}(1993)\citenamefont
  {Halperin}, \citenamefont {Lee},\ and\ \citenamefont {Read}}]{HLR}%
  \BibitemOpen
  \bibfield  {author} {\bibinfo {author} {\bibfnamefont {B.~I.}\ \bibnamefont
  {Halperin}}, \bibinfo {author} {\bibfnamefont {P.~A.}\ \bibnamefont {Lee}}, \
  and\ \bibinfo {author} {\bibfnamefont {N.}~\bibnamefont {Read}},\ }\href
  {\doibase 10.1103/PhysRevB.47.7312} {\bibfield  {journal} {\bibinfo
  {journal} {Phys. Rev. B}\ }\textbf {\bibinfo {volume} {47}},\ \bibinfo
  {pages} {7312} (\bibinfo {year} {1993})}\BibitemShut {NoStop}%
\bibitem [{\citenamefont {Jain}(1989)}]{Jain}%
  \BibitemOpen
  \bibfield  {author} {\bibinfo {author} {\bibfnamefont {J.~K.}\ \bibnamefont
  {Jain}},\ }\href {\doibase 10.1103/PhysRevLett.63.199} {\bibfield  {journal}
  {\bibinfo  {journal} {Phys. Rev. Lett.}\ }\textbf {\bibinfo {volume} {63}},\
  \bibinfo {pages} {199} (\bibinfo {year} {1989})}\BibitemShut {NoStop}%
\bibitem [{\citenamefont {Dasgupta}\ and\ \citenamefont
  {Halperin}(1981)}]{Dasgupta}%
  \BibitemOpen
  \bibfield  {author} {\bibinfo {author} {\bibfnamefont {C.}~\bibnamefont
  {Dasgupta}}\ and\ \bibinfo {author} {\bibfnamefont {B.~I.}\ \bibnamefont
  {Halperin}},\ }\href {\doibase 10.1103/PhysRevLett.47.1556} {\bibfield
  {journal} {\bibinfo  {journal} {Phys. Rev. Lett.}\ }\textbf {\bibinfo
  {volume} {47}},\ \bibinfo {pages} {1556} (\bibinfo {year}
  {1981})}\BibitemShut {NoStop}%
\bibitem [{\citenamefont {Fisher}\ and\ \citenamefont {Lee}(1989)}]{Fisher}%
  \BibitemOpen
  \bibfield  {author} {\bibinfo {author} {\bibfnamefont {M.~P.~A.}\
  \bibnamefont {Fisher}}\ and\ \bibinfo {author} {\bibfnamefont {D.~H.}\
  \bibnamefont {Lee}},\ }\href {\doibase 10.1103/PhysRevB.39.2756} {\bibfield
  {journal} {\bibinfo  {journal} {Phys. Rev. B}\ }\textbf {\bibinfo {volume}
  {39}},\ \bibinfo {pages} {2756} (\bibinfo {year} {1989})}\BibitemShut
  {NoStop}%
\bibitem [{\citenamefont {Mross}\ \emph {et~al.}(2015)\citenamefont {Mross},
  \citenamefont {Essin},\ and\ \citenamefont {Alicea}}]{Mross2015}%
  \BibitemOpen
  \bibfield  {author} {\bibinfo {author} {\bibfnamefont {D.~F.}\ \bibnamefont
  {Mross}}, \bibinfo {author} {\bibfnamefont {A.}~\bibnamefont {Essin}}, \ and\
  \bibinfo {author} {\bibfnamefont {J.}~\bibnamefont {Alicea}},\ }\href
  {\doibase 10.1103/PhysRevX.5.011011} {\bibfield  {journal} {\bibinfo
  {journal} {Phys. Rev. X}\ }\textbf {\bibinfo {volume} {5}},\ \bibinfo {pages}
  {011011} (\bibinfo {year} {2015})}\BibitemShut {NoStop}%
\bibitem [{\citenamefont {{Thanh Son}}(2015)}]{Son}%
  \BibitemOpen
  \bibfield  {author} {\bibinfo {author} {\bibfnamefont {D.}~\bibnamefont
  {{Thanh Son}}},\ }\href@noop {} {\bibfield  {journal} {\bibinfo  {journal}
  {ArXiv e-prints}\ } (\bibinfo {year} {2015})},\ \Eprint
  {http://arxiv.org/abs/1502.03446} {arXiv:1502.03446 [cond-mat.mes-hall]}
  \BibitemShut {NoStop}%
\bibitem [{\citenamefont {{Barkeshli}}\ \emph {et~al.}(2015)\citenamefont
  {{Barkeshli}}, \citenamefont {{Mulligan}},\ and\ \citenamefont
  {{Fisher}}}]{Barkeshli}%
  \BibitemOpen
  \bibfield  {author} {\bibinfo {author} {\bibfnamefont {M.}~\bibnamefont
  {{Barkeshli}}}, \bibinfo {author} {\bibfnamefont {M.}~\bibnamefont
  {{Mulligan}}}, \ and\ \bibinfo {author} {\bibfnamefont {M.~P.~A.}\
  \bibnamefont {{Fisher}}},\ }\href@noop {} {\bibfield  {journal} {\bibinfo
  {journal} {ArXiv e-prints}\ } (\bibinfo {year} {2015})},\ \Eprint
  {http://arxiv.org/abs/1502.05404} {arXiv:1502.05404 [cond-mat.str-el]}
  \BibitemShut {NoStop}%
\bibitem [{\citenamefont {{Kitaev}}(2009)}]{KitaevNI}%
  \BibitemOpen
  \bibfield  {author} {\bibinfo {author} {\bibfnamefont {A.}~\bibnamefont
  {{Kitaev}}},\ }in\ \href {\doibase 10.1063/1.3149495} {\emph {\bibinfo
  {booktitle} {American Institute of Physics Conference Series}}},\ \bibinfo
  {series} {American Institute of Physics Conference Series}, Vol.\ \bibinfo
  {volume} {1134},\ \bibinfo {editor} {edited by\ \bibinfo {editor}
  {\bibfnamefont {V.}~\bibnamefont {{Lebedev}}}\ and\ \bibinfo {editor}
  {\bibfnamefont {M.}~\bibnamefont {{Feigel'Man}}}}\ (\bibinfo {year} {2009})\
  pp.\ \bibinfo {pages} {22--30},\ \Eprint {http://arxiv.org/abs/0901.2686}
  {arXiv:0901.2686 [cond-mat.mes-hall]} \BibitemShut {NoStop}%
\bibitem [{\citenamefont {Schnyder}\ \emph {et~al.}(2008)\citenamefont
  {Schnyder}, \citenamefont {Ryu}, \citenamefont {Furusaki},\ and\
  \citenamefont {Ludwig}}]{LudwigNI}%
  \BibitemOpen
  \bibfield  {author} {\bibinfo {author} {\bibfnamefont {A.~P.}\ \bibnamefont
  {Schnyder}}, \bibinfo {author} {\bibfnamefont {S.}~\bibnamefont {Ryu}},
  \bibinfo {author} {\bibfnamefont {A.}~\bibnamefont {Furusaki}}, \ and\
  \bibinfo {author} {\bibfnamefont {A.~W.~W.}\ \bibnamefont {Ludwig}},\ }\href
  {\doibase 10.1103/PhysRevB.78.195125} {\bibfield  {journal} {\bibinfo
  {journal} {Phys. Rev. B}\ }\textbf {\bibinfo {volume} {78}},\ \bibinfo
  {pages} {195125} (\bibinfo {year} {2008})}\BibitemShut {NoStop}%
\bibitem [{\citenamefont {{Witten}}(1979)}]{WittenEffect}%
  \BibitemOpen
  \bibfield  {author} {\bibinfo {author} {\bibfnamefont {E.}~\bibnamefont
  {{Witten}}},\ }\href {\doibase 10.1016/0370-2693(79)90838-4} {\bibfield
  {journal} {\bibinfo  {journal} {Physics Letters B}\ }\textbf {\bibinfo
  {volume} {86}},\ \bibinfo {pages} {283} (\bibinfo {year} {1979})}\BibitemShut
  {NoStop}%
\bibitem [{\citenamefont {Metlitski}()}]{Metlitski_ta}%
  \BibitemOpen
  \bibfield  {author} {\bibinfo {author} {\bibfnamefont {M.}~\bibnamefont
  {Metlitski}},\ }\href@noop {} {}\bibinfo {howpublished} {to
  appear}\BibitemShut {NoStop}%
\bibitem [{\citenamefont {Vadim~Borokhov}(2002)}]{KapustinMonopoles}%
  \BibitemOpen
  \bibfield  {author} {\bibinfo {author} {\bibfnamefont {X.~W.}\ \bibnamefont
  {Vadim~Borokhov}, \bibfnamefont {Anton~Kapustin}},\ }\href@noop {} {\bibfield
   {journal} {\bibinfo  {journal} {JHEP}\ }\textbf {\bibinfo {volume} {11}},\
  \bibinfo {pages} {49} (\bibinfo {year} {2002})}\BibitemShut {NoStop}%
\bibitem [{\citenamefont {Kapustin}(2014)}]{KapustinBos}%
  \BibitemOpen
  \bibfield  {author} {\bibinfo {author} {\bibfnamefont {A.}~\bibnamefont
  {Kapustin}},\ }\href@noop {} {\bibfield  {journal} {\bibinfo  {journal}
  {ArXiv e-prints: 1403.1467}\ } (\bibinfo {year} {2014})}\BibitemShut
  {NoStop}%
\bibitem [{\citenamefont {Hermele}\ \emph {et~al.}(2005)\citenamefont
  {Hermele}, \citenamefont {Senthil},\ and\ \citenamefont
  {Fisher}}]{HermeleFisher}%
  \BibitemOpen
  \bibfield  {author} {\bibinfo {author} {\bibfnamefont {M.}~\bibnamefont
  {Hermele}}, \bibinfo {author} {\bibfnamefont {T.}~\bibnamefont {Senthil}}, \
  and\ \bibinfo {author} {\bibfnamefont {M.~P.~A.}\ \bibnamefont {Fisher}},\
  }\href {\doibase 10.1103/PhysRevB.72.104404} {\bibfield  {journal} {\bibinfo
  {journal} {Phys. Rev. B}\ }\textbf {\bibinfo {volume} {72}},\ \bibinfo
  {pages} {104404} (\bibinfo {year} {2005})}\BibitemShut {NoStop}%
\bibitem [{\citenamefont {Anton~Kapustin}\ and\ \citenamefont
  {Wang}(2014)}]{KapustinFerm}%
  \BibitemOpen
  \bibfield  {author} {\bibinfo {author} {\bibfnamefont {A.~T.}\ \bibnamefont
  {Anton~Kapustin}, \bibfnamefont {Ryan~Thorngren}}\ and\ \bibinfo {author}
  {\bibfnamefont {Z.}~\bibnamefont {Wang}},\ }\href@noop {} {\bibfield
  {journal} {\bibinfo  {journal} {ArXiv eprints 1406.7329}\ } (\bibinfo {year}
  {2014})}\BibitemShut {NoStop}%
\bibitem [{\citenamefont {Walker}\ and\ \citenamefont {Wang}(2012)}]{Walker12}%
  \BibitemOpen
  \bibfield  {author} {\bibinfo {author} {\bibfnamefont {K.}~\bibnamefont
  {Walker}}\ and\ \bibinfo {author} {\bibfnamefont {Z.}~\bibnamefont {Wang}},\
  }\href@noop {} {\bibfield  {journal} {\bibinfo  {journal} {Front. Phys.}\
  }\textbf {\bibinfo {volume} {7}},\ \bibinfo {pages} {150} (\bibinfo {year}
  {2012})}\BibitemShut {NoStop}%
\bibitem [{\citenamefont {{Rychkov}}(2011)}]{Rychkov}%
  \BibitemOpen
  \bibfield  {author} {\bibinfo {author} {\bibfnamefont {S.}~\bibnamefont
  {{Rychkov}}},\ }\href@noop {} {\bibfield  {journal} {\bibinfo  {journal}
  {ArXiv e-prints}\ } (\bibinfo {year} {2011})},\ \Eprint
  {http://arxiv.org/abs/1111.2115} {arXiv:1111.2115 [hep-th]} \BibitemShut
  {NoStop}%
\bibitem [{\citenamefont {Grover}\ \emph {et~al.}(2014)\citenamefont {Grover},
  \citenamefont {Sheng},\ and\ \citenamefont {Vishwanath}}]{Grover}%
  \BibitemOpen
  \bibfield  {author} {\bibinfo {author} {\bibfnamefont {T.}~\bibnamefont
  {Grover}}, \bibinfo {author} {\bibfnamefont {D.~N.}\ \bibnamefont {Sheng}}, \
  and\ \bibinfo {author} {\bibfnamefont {A.}~\bibnamefont {Vishwanath}},\
  }\href {\doibase 10.1126/science.1248253} {\bibfield  {journal} {\bibinfo
  {journal} {Science}\ }\textbf {\bibinfo {volume} {344}},\ \bibinfo {pages}
  {280} (\bibinfo {year} {2014})}\BibitemShut {NoStop}%
\bibitem [{\citenamefont {{Ponte}}\ and\ \citenamefont
  {{Lee}}(2014)}]{PonteLee}%
  \BibitemOpen
  \bibfield  {author} {\bibinfo {author} {\bibfnamefont {P.}~\bibnamefont
  {{Ponte}}}\ and\ \bibinfo {author} {\bibfnamefont {S.-S.}\ \bibnamefont
  {{Lee}}},\ }\href {\doibase 10.1088/1367-2630/16/1/013044} {\bibfield
  {journal} {\bibinfo  {journal} {New Journal of Physics}\ }\textbf {\bibinfo
  {volume} {16}},\ \bibinfo {eid} {013044} (\bibinfo {year} {2014})},\ \Eprint
  {http://arxiv.org/abs/1206.2340} {arXiv:1206.2340 [cond-mat.str-el]}
  \BibitemShut {NoStop}%
\bibitem [{\citenamefont {Aharony}\ \emph {et~al.}(1997)\citenamefont
  {Aharony}, \citenamefont {Hanany}, \citenamefont {Intriligator},
  \citenamefont {Seiberg},\ and\ \citenamefont {Strassler}}]{Aharony}%
  \BibitemOpen
  \bibfield  {author} {\bibinfo {author} {\bibfnamefont {O.}~\bibnamefont
  {Aharony}}, \bibinfo {author} {\bibfnamefont {A.}~\bibnamefont {Hanany}},
  \bibinfo {author} {\bibfnamefont {K.~A.}\ \bibnamefont {Intriligator}},
  \bibinfo {author} {\bibfnamefont {N.}~\bibnamefont {Seiberg}}, \ and\
  \bibinfo {author} {\bibfnamefont {M.}~\bibnamefont {Strassler}},\ }\href
  {\doibase 10.1016/S0550-3213(97)00323-4} {\bibfield  {journal} {\bibinfo
  {journal} {Nucl.Phys.}\ }\textbf {\bibinfo {volume} {B499}},\ \bibinfo
  {pages} {67} (\bibinfo {year} {1997})},\ \Eprint
  {http://arxiv.org/abs/hep-th/9703110} {arXiv:hep-th/9703110 [hep-th]}
  \BibitemShut {NoStop}%
\bibitem [{\citenamefont {{Borokhov}}\ \emph {et~al.}(2002)\citenamefont
  {{Borokhov}}, \citenamefont {{Kapustin}},\ and\ \citenamefont
  {{Wu}}}]{KapustinBorokhov}%
  \BibitemOpen
  \bibfield  {author} {\bibinfo {author} {\bibfnamefont {V.}~\bibnamefont
  {{Borokhov}}}, \bibinfo {author} {\bibfnamefont {A.}~\bibnamefont
  {{Kapustin}}}, \ and\ \bibinfo {author} {\bibfnamefont {X.}~\bibnamefont
  {{Wu}}},\ }\href {\doibase 10.1088/1126-6708/2002/12/044} {\bibfield
  {journal} {\bibinfo  {journal} {Journal of High Energy Physics}\ }\textbf
  {\bibinfo {volume} {12}},\ \bibinfo {eid} {044} (\bibinfo {year} {2002})},\
  \Eprint {http://arxiv.org/abs/hep-th/0207074} {hep-th/0207074} \BibitemShut
  {NoStop}%
\bibitem [{\citenamefont {{Wang}}\ and\ \citenamefont
  {{Senthil}}(2015)}]{Wang2015}%
  \BibitemOpen
  \bibfield  {author} {\bibinfo {author} {\bibfnamefont {C.}~\bibnamefont
  {{Wang}}}\ and\ \bibinfo {author} {\bibfnamefont {T.}~\bibnamefont
  {{Senthil}}},\ }\href@noop {} {\bibfield  {journal} {\bibinfo  {journal}
  {ArXiv e-prints}\ } (\bibinfo {year} {2015})},\ \Eprint
  {http://arxiv.org/abs/1505.03520} {arXiv:1505.03520 [cond-mat.str-el]}
  \BibitemShut {NoStop}%
\end{thebibliography}%

\end{document}